\documentclass{JINST}

\usepackage{amsmath}
\usepackage{amssymb}
\usepackage{xspace}
\usepackage{textcomp}
\usepackage{setspace}
\usepackage{booktabs}
\usepackage{lineno}

\newcommand{\be}{\begin{equation}}
\newcommand{\ee}{\end{equation}}
\newcommand{\ben}{\begin{enumerate}}
\newcommand{\een}{\end{enumerate}}
\newcommand{\bi}{\begin{itemize}}
\newcommand{\ei}{\end{itemize}}
\newcommand{\bbe}{\begin{equation*}}
\newcommand{\eee}{\end{equation*}}
\newcommand{\bber}{\begin{equation*}\textcolor{red}}
\newcommand{\eeer}{\end{equation*}}
\newcommand{\bc}{\begin{center}}
\newcommand{\ec}{\end{center}}
\newcommand{\bea}{\begin{eqnarray}}
\newcommand{\eea}{\end{eqnarray}}
\newcommand{\bem}{\begin{pmatrix}}
\newcommand{\eem}{\end{pmatrix}}
\newcommand{\bbea}{\begin{eqnarray*}}
\newcommand{\eeea}{\end{eqnarray*}}

\newcommand{\bts}{{\emph {BtS}}\xspace}
\newcommand{\sts}{{\emph {StS}}\xspace}
\newcommand{\autoscan}{{\emph {AutoScan}}\xspace}
\newcommand{\nmmm}{{\emph {nMMM}}\xspace}
\newcommand{\sbts}{S$_{\text{BtS}}$\xspace}
\newcommand{\snmmm}{S$_{\text{nMMM}}$\xspace}
\newcommand{\sgdas}{S$_{\text{GDAS}}$\xspace}
\newcommand{\mr}[1]{\mathrm{#1}}

\newcommand{\xmax}{{$X_\text{max}$}\xspace}
\newcommand{\xmaxOne}{{$X_\text{max,1}$}\xspace}
\newcommand{\xmaxTwo}{{$X_\text{max,2}$}\xspace}
\def\EeV{\ifmmode {\mathrm{\ Ee\kern -0.1em V}}\else
                   \textrm{Ee\kern -0.1em V}\fi\xspace}%
\def\PeV{\ifmmode {\mathrm{\ Pe\kern -0.1em V}}\else
                   \textrm{Pe\kern -0.1em V}\fi\xspace}%
\def\TeV{\ifmmode {\mathrm{\ Te\kern -0.1em V}}\else
                   \textrm{Te\kern -0.1em V}\fi\xspace}%
\def\MeV{\ifmmode {\mathrm{\ Me\kern -0.1em V}}\else
                   \textrm{Me\kern -0.1em V}\fi\xspace}%
\def\GeV{\ifmmode {\mathrm{\ Ge\kern -0.1em V}}\else
                   \textrm{Ge\kern -0.1em V}\fi\xspace}%
\def\keV{\ifmmode {\mathrm{\ ke\kern -0.1em V}}\else
                   \textrm{ke\kern -0.1em V}\fi\xspace}%
\def\MeV{\ifmmode {\mathrm{\ Me\kern -0.1em V}}\else
                   \textrm{Me\kern -0.1em V}\fi\xspace}%
\def\eV{\ifmmode {\mathrm{\ e\kern -0.1em V}}\else
                   \textrm{e\kern -0.1em V}\fi\xspace}%

\def \mal      {Ma\-lar\-g\"{u}e\xspace}
\def \pao      {Pierre Auger Observatory\xspace}
\def \relxmax  {$\Delta X_\text{max}$\xspace}
\def \relEavg  {$\Delta E/\langle E\rangle$\xspace}
\def \gcmsq    {$\text{g~cm}^{-2}$\xspace}
\def\Offline{\mbox{$\overline{\textrm%
{Off}}$\hspace{.05em}\protect\raisebox{.4ex}%
{$\protect\underline{\textrm{line}}$}}\xspace}

\title{The Rapid Atmospheric Monitoring System of the\\ \pao}
\author{The Pierre Auger Collaboration\footnote{Authors are listed on the
following pages. E-mail: auger\_spokespersons@fnal.gov}}

\abstract
{
  The \pao is a facility built to detect air showers produced by cosmic rays
  above $10^{17}$~eV.  During clear nights with a low illuminated moon
  fraction, the UV fluorescence light produced by air showers is recorded by
  optical telescopes at the Observatory.  To correct the observations for
  variations in atmospheric conditions, atmospheric monitoring is performed at
  regular intervals ranging from several minutes (for cloud identification) to
  several hours (for aerosol conditions) to several days (for vertical profiles
  of temperature, pressure, and humidity).  In 2009, the monitoring program was
  upgraded to allow for additional targeted measurements of atmospheric
  conditions shortly after the detection of air showers of special interest,
  e.\,g., showers produced by very high-energy cosmic rays or showers with
  atypical longitudinal profiles.  The former events are of particular
  importance for the determination of the energy scale of the Observatory, and
  the latter are characteristic of unusual air shower physics or exotic primary
  particle types.  The purpose of targeted (or ``rapid'') monitoring is to
  improve the resolution of the atmospheric measurements for such events.  In
  this paper, we report on the implementation of the rapid monitoring program
  and its current status.  The rapid monitoring data have been analyzed and
  applied to the reconstruction of air showers of high interest, and indicate
  that the air fluorescence measurements affected by clouds and aerosols are
  effectively corrected using measurements from the regular atmospheric
  monitoring program.  We find that the rapid monitoring program has potential
  for supporting dedicated physics analyses beyond the standard event
  reconstruction.
}

\keywords{Cosmic rays, extensive air showers, air fluorescence method,
atmospheric monitoring, calibration, radiosonde, lidar, star monitoring}

\begin{document}
\setlength{\topmargin}{3mm}

\newpage
% This is a fragment that can be inserted into a LaTeX
\par\noindent
{\bf The Pierre Auger Collaboration} \\
P.~Abreu$^{63}$, 
M.~Aglietta$^{51}$, 
M.~Ahlers$^{94}$, 
E.J.~Ahn$^{81}$, 
I.F.M.~Albuquerque$^{15}$, 
D.~Allard$^{29}$, 
I.~Allekotte$^{1}$, 
J.~Allen$^{85}$, 
P.~Allison$^{87}$, 
A.~Almela$^{11,\: 7}$
% ...

\noindent
{\it\footnotesize
$^{1}$ Centro At\'{o}mico Bariloche and Instituto Balseiro (CNEA-UNCuyo-CONICET), San 
Carlos de Bariloche, 
Argentina \\
% ...
$^{7}$ Instituto de Tecnolog\'{\i}as en Detecci\'{o}n y Astropart\'{\i}culas (CNEA, CONICET, UNSAM), 
Buenos Aires, 
Argentina \\
% ...
$^{11}$ Universidad Tecnol\'{o}gica Nacional - Facultad Regional Buenos Aires, Buenos Aires,
Argentina \\
% ...
$^{15}$ Universidade de S\~{a}o Paulo, Instituto de F\'{\i}sica, S\~{a}o Paulo, SP, 
Brazil \\
% ...
$^{29}$ Laboratoire AstroParticule et Cosmologie (APC), Universit\'{e} Paris 7, CNRS-IN2P3, 
Paris, 
France \\
% ...
$^{51}$ Istituto di Fisica dello Spazio Interplanetario (INAF), Universit\`{a} di Torino and 
Sezione INFN, Torino, 
Italy \\
% ...
$^{63}$ LIP and Instituto Superior T\'{e}cnico, Technical University of Lisbon, 
Portugal \\
% ...
$^{81}$ Fermilab, Batavia, IL, 
USA \\
% ...
$^{85}$ New York University, New York, NY, 
USA \\
% ...
$^{87}$ Ohio State University, Columbus, OH, 
USA \\
% ...
$^{94}$ University of Wisconsin, Madison, WI, 
USA \\
% ...
\par\noindent
(\ddag) Deceased \\
(a) at Konan University, Kobe, Japan \\
(b) now at the Universidad Autonoma de Chiapas on leave of absence from Cinvestav \\
(f) now at University of Maryland \\
(h) now at NYU Abu Dhabi \\
(i) now at Universit\'{e} de Lausanne \\
}
% last updated:	6/4/2012 

\newpage

\section{Introduction}\label{sec:introduction}

The \pao, located about 1\,400~meters above sea level near the town of \mal,
Argentina, is designed to observe extensive air showers created by cosmic rays with
energies above $10^{17}$~eV.  Multiple complementary air shower detectors are
operated at the Observatory to overcome the shortcomings of any single
measurement technique.

The primary instrument of the \pao is a large-area Surface Detector
(SD)~\cite{Allekotte:2007sf,Suomijarvi:2009icrc}, which is used to sample
the secondary particles from air showers that reach the ground.  The SD is an
array of about 1\,600 water Cherenkov stations arranged $1.5$~km apart on a
triangular grid.  The array is deployed over an area of 3\,000~km$^2$, and it
has a duty cycle of nearly 100\%.  Thus, data from the SD provide a
high-statistics sample of air showers used to study the energy
spectrum and arrival direction distribution of the cosmic rays above
$10^{17}$~eV.

While the SD is sensitive to the lateral distribution of secondary air shower
particles  at ground level, the longitudinal development of showers in the
atmosphere is measured using a Fluorescence Detector (FD) of 27 optical
telescopes~\cite{Abraham:2009pm}. The telescopes, optimized for the
near-ultraviolet band, are located at four sites on the periphery of the SD
array: Los Leones, Los Morados, Loma Amarilla, and Coihueco (see
Fig.~\ref{fig:moniSystems}). Each site is instrumented with six telescopes
deployed inside a climate-controlled building. Together the six telescopes have
a field of view covering 180$^\circ$ in azimuth and about 0$^\circ$ to
30$^\circ$ in elevation. At Coihueco, three additional High-Elevation Auger
Telescopes (HEAT) have been deployed to observe elevation angles between
30$^\circ$ and 60$^\circ$ \cite{Meurer:2011ms}.

The fluorescence telescopes are capable of recording the ultraviolet
fluorescence and Cherenkov light produced during air shower development.  The
flux of fluorescence photons from a given point in an air shower track is
proportional to d$E$/d$X$, the energy loss per unit slant depth $X$ of
traversed atmosphere~\cite{Arqueros:2008}.  The Cherenkov emission is
proportional to the number of charged particles in the shower above the
Cherenkov production threshold, and depends on the energy loss and energy
distribution of secondary electrons and positrons in the shower.  By observing
the UV emission from an air shower, it is possible to observe the energy loss
as a function of $X$ and make a calorimetric estimate of the energy of the
primary particle, after correcting for ``missing energy'' not contained in the
electromagnetic component of the shower~\cite{Unger:2008uq}.  The slant depth
at which the energy deposition rate d$E$/d$X$ reaches its maximum value is
called \xmax.  By observing \xmax for a large set of air showers, the FD data
can be used to discuss the composition and the interaction properties of cosmic
rays as a function of primary energy~\cite{Abraham:2010yv}.

Simultaneous measurements of air showers with the FD and SD are called hybrid
events.  By performing a joint reconstruction which uses geometrical and
timing information from both detectors, it is possible to significantly improve
the angular and energy resolution of reconstructed hybrid events with respect
to showers observed by the FD alone~\cite{Mostafa:2006id}. Therefore, when FD
data are used to produce physics results, only hybrid events are included in the
analysis.  Moreover, events observed with high quality in hybrid mode are
crucial for the calibration of measurements performed using the SD.  While the
energy of a primary cosmic ray can be estimated using data from the SD alone,
the absolute scale of the energy estimator depends on hadronic interaction
models of air shower development.  To remove this model dependence, the energy
scale of the SD is calibrated using a subsample of the hybrid events in which a
calorimetric energy measurement from the FD can be compared to an independent
energy estimate from the SD~\cite{Pesce:2011icrc}.

The FD is only operated during nights when UV light from air showers is not
overwhelmed by moonlight.  Safe telescope operations also require adequate
weather conditions (i.\,e., no rain and moderate wind) and high atmospheric
transmittance to assure high data quality.  These restrictions limit the duty cycle
of the FD to about $12\%$~\cite{Salamida:2011}.  As a result, at trigger level
the number of events observed with the FD is an order of magnitude smaller than
that observed with the SD.

The light profiles recorded with the fluorescence telescopes must be corrected
for UV attenuation along light paths of up to $40$~km.  To estimate the
attenuation of light by molecules, aerosols, and clouds, regular atmospheric
measurements are performed at the Observatory using UV laser shots, radiosonde
launches, optical observations, and cloud measurements in the
mid-infrared~\cite{Abraham:2010}.  The radiosondes provide measurements of the
main atmospheric state variables such as temperature, pressure, and humidity,
which affect mainly the production of fluorescence light induced by air
showers~\cite{Arqueros:2008,Keilhauer:2009icrc2}, but also the light scattering
by molecules.  The laser shots and optical observations are used to estimate
the aerosol optical depth and the cloud cover over the FD buildings.

The regular atmospheric monitoring performed at the Observatory provides
atmospheric data of local conditions with a time resolution of several minutes
to several days, depending on the type of measurement. This is sufficient for
the bulk of measured air showers.  Hourly and daily atmospheric corrections are
available for reconstructing individual showers, and the average energy
dependence of the atmospheric corrections for the full sample of observed
cosmic rays is well-understood~\cite{Abraham:2010}.  However, because of the
massive volume of atmosphere used to perform fluorescence observations
--~nearly 30\,000~km$^3$~-- the time and spatial resolution of the atmospheric
database is necessarily limited.

For some analyses, it is desirable to provide atmospheric data beyond the
regular measurements.  For example, the high-energy tail of the data sample
used in the SD energy calibration is an important lever arm in the SD-FD fit.
Since atmospheric corrections are of utmost importance for the highest-energy
showers recorded with the FD, it is sensible to perform dedicated atmospheric
measurements at the time and location of high-energy cosmic ray events.  Other
showers of interest are anomalous longitudinal profiles
observed in the FD data.  The rate of these showers is expected to be largest
at low energies and for light primary masses~\cite{Baus:2011icrc}. Such showers
are removed by standard analysis cuts because lumpy profiles are typically
caused by atmospheric non-uniformities such as cloud banks or aerosol layers.
However, these profiles may also be indicators of exotic primary particles or
unusual air shower development.  In any analysis which uses longitudinal
profiles to search for such exotic phenomena, dedicated monitoring of
air-shower tracks is needed to remove events which could be distorted by
atmospheric effects.

To provide high-resolution atmospheric data for interesting air
showers, we have implemented an automatic online monitoring system which can be
used to trigger dedicated atmospheric measurements a few minutes after the air
showers are detected.  This rapid monitoring trigger was commissioned in early
2009 and has been integrated into the regular monitoring schedules of several of
the atmospheric monitoring subsystems.  In this paper, we will discuss the
operation and performance of the rapid monitoring program.  In
Section~\ref{sec:fl_meas}, we describe the \pao and review the standard
atmospheric monitoring program.  The implementation of the online atmospheric
monitor is discussed in Section~\ref{sec:implementation}.  The integration of
rapid monitoring into the radiosonde, lidar, and optical telescope subsystems is
discussed in Sections~\ref{sec:bts}, \ref{sec:sts}, and \ref{sec:fram}, along
with a selection of interesting showers. We conclude in
Section~\ref{sec:conclusion}.

\section{Atmospheric Monitoring
\label{sec:fl_meas}}

\begin{figure}[htb]
  \begin{center}
    \includegraphics*[width=0.5\textwidth,clip]{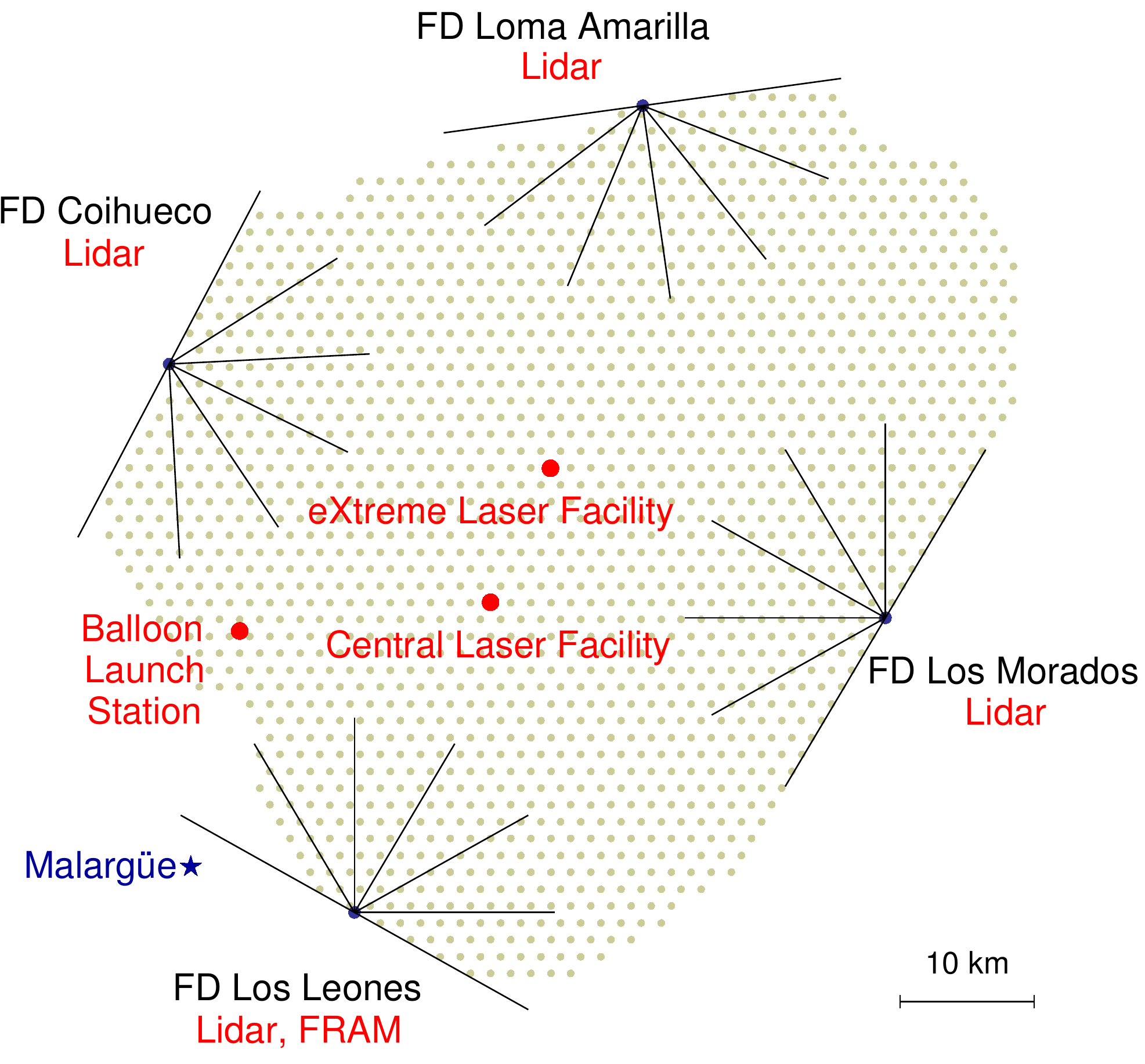}
    \caption[]{\label{fig:moniSystems} \slshape
      Layout of the surface detector array (dotted area) and fluorescence
      telescope sites, showing only the locations of the atmospheric monitoring
      subsystems which are integrated into the rapid monitoring program.  The
      two central laser facilities, which do not receive rapid monitoring
      triggers, are shown for reference.
    }
  \end{center}
\end{figure}

As described in Section~\ref{sec:introduction}, measurements of air showers with
the fluorescence telescopes are affected by fluctuations in atmospheric
conditions, and so extensive atmospheric monitoring is carried out at the
Observatory~\cite{Abraham:2010}.  The locations of the SD, FD, and the
atmospheric monitors described in this work are shown in
Fig.~\ref{fig:moniSystems}.

Atmospheric measurements are stored in several multi-gigabyte databases for use
in the offline reconstruction of air showers.  The time resolution of the
measurements ranges between five minutes (in the case of cloud data) to one hour
(in the case of aerosol data) to several days (in the case of altitude-dependent
atmospheric state variables).  The spatial resolution is limited, the
altitude-dependent atmospheric state variables are assumed to be horizontally
uniform across the SD array, while aerosol conditions and state variables from
ground-based weather stations are treated as uniform in the region around each
FD building or station, respectively.  The systematic uncertainties introduced
by the limited resolution of the database have been estimated and are reported
as part of the uncertainty in the FD energy scale provided for the SD energy
calibration~\cite{Abraham:2010,Dawson:2011zz}.  Due to the correlation between
the reconstructed energies of air showers and the distances at which they are
observed in the telescopes, the uncertainties increase linearly with
energy~\cite{Abraham:2010}.

\subsection{Atmospheric State Variables and Site Models
\label{subsec:nMMM}}

\begin{figure}[tbp]
  \centering
  \includegraphics[width=1.\textwidth,clip]{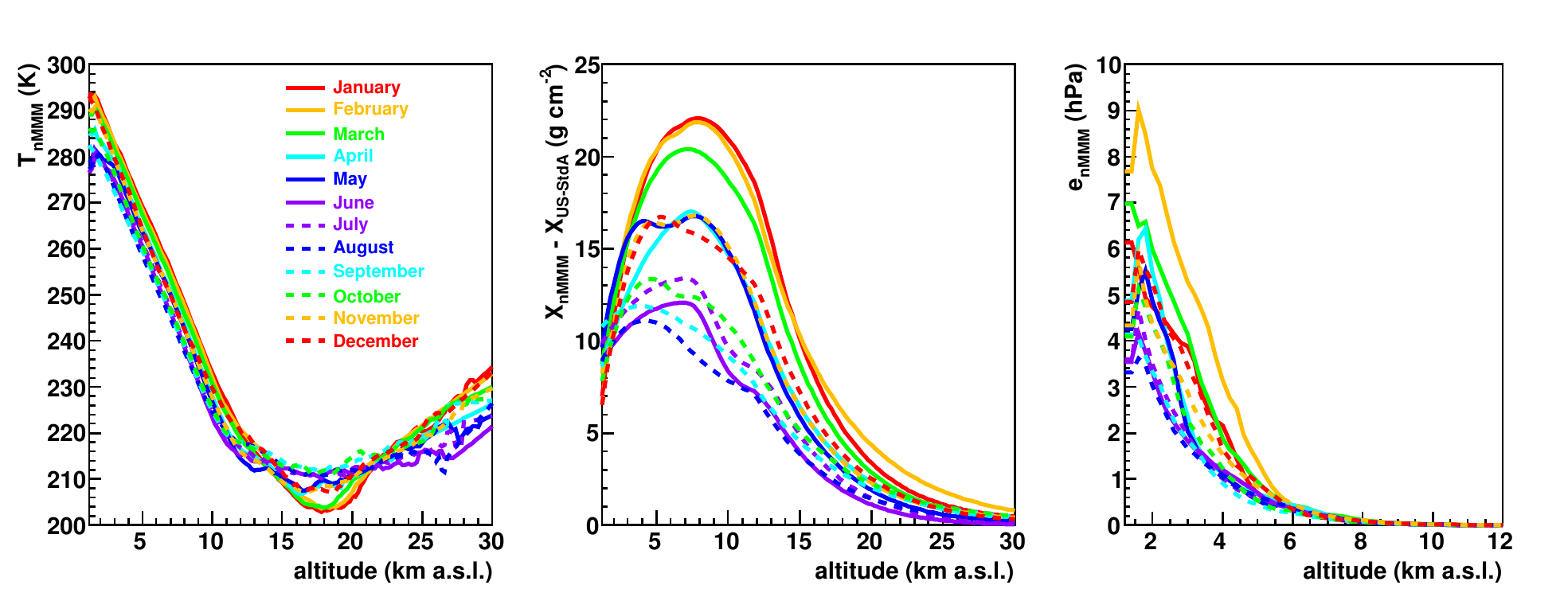}
  \includegraphics[width=1.\textwidth,clip]{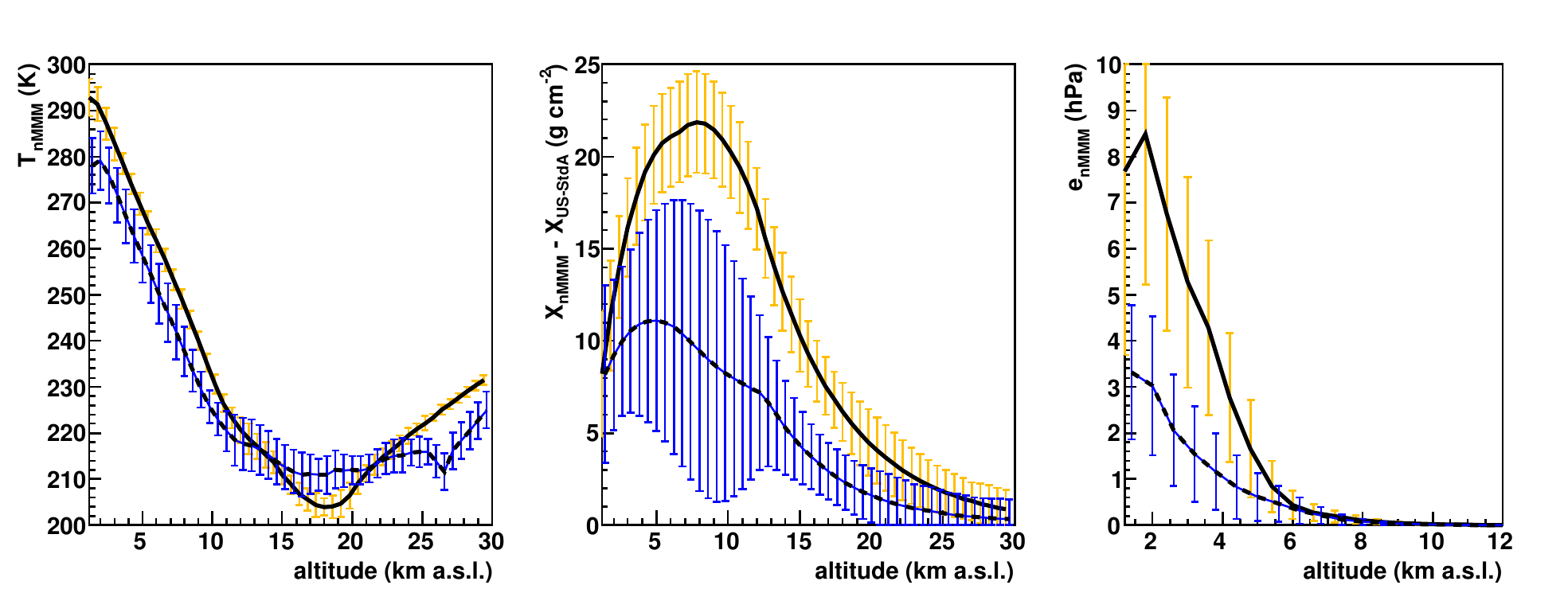}
  \caption[]{\label{fig:nmmm} \slshape
    Top: Atmospheric profiles of the \emph{new \mal Monthly Model}, left:
    temperature, middle: atmospheric depth, right: water vapor pressure.  The
    depth profiles are expressed with respect to the U.S. Standard
    Atmosphere~\cite{USStdAtm:1976}.
    Bottom: The same graphs as in upper row with uncertainties for February
    (solid line with yellow uncertainties) and August (dashed line
    with blue uncertainties).}
\end{figure}

Air temperature, pressure, wind speed, and humidity are recorded at ground level
by weather stations at each FD building and at the Central Laser Facility (see
Fig.~\ref{fig:moniSystems}), and between 2002 and 2010 a weather balloon program
was operated at the \pao.  Prior to mid-2005, the radio soundings were performed
in ten dedicated campaigns, each lasting two to three weeks, with an average of
10~launches per campaign.  Between mid-2005 and end of 2008, the balloon
launches were performed more regularly -- about every five days and
independently of FD data-taking.

To compensate for the missing information between the radiosonde measurements,
average models of monthly conditions were constructed. The first version of
these \mal Monthly Models (\textsl{MMM}) contained vertical profiles of
atmospheric temperature $T$, pressure $p$, density $\rho$, and atmospheric depth
$X$ derived from pre-2005 weather data from \mal and data from C\'{o}rdoba and
Santa Rosa, Argentina, the sites nearest \mal with publicly available radio
sounding measurements~\cite{Keilhauer:2005ja}. The local measurements were
supplemented with external data because of the low measurement statistics at the
Observatory when the models were constructed.  By 2009, the number of balloon
flights over the Observatory was sufficient to re-evaluate the profiles and
construct improved models with an additional average profile of the water vapor
pressure $e$~\cite{Abraham:2010}.  These new \mal Monthly Models (\nmmm) were
derived from 261~local radio soundings performed between August 2002 and
December 2008.

The \nmmm profiles comprise vertical profiles of $T$, $p$, $\rho$, $X$, and $e$
specified between $1.2$~km and $30$~km above sea level in steps of $200$~m.  Of
the 261~radio soundings used to construct the models, 32 were discarded during
construction of the vapor pressure profiles due to contamination of the balloon
flights by high cloud coverage.  Above $12$~km, the vapor pressure has been set
to zero.

The local radio soundings provide reliable and unbiased measurements of the
monthly average profiles between about $1.6$~km and the burst altitude of the
balloons.  The burst altitude was typically at $23$~km, with a few balloons
reaching a maximum altitude of $27$~km.  Data from the five ground-based weather
stations at the Observatory were used to extrapolate the profiles down to
$1.2$~km\footnote{For technical reasons during air shower reconstruction, the
profiles need to go beyond the lowest surface height.}.  Above the altitude of
balloon burst, the data have been extrapolated using values from the 2005
monthly models.  The \nmmm profiles of $T$, $X$, and $e$ are shown in
Fig.~\ref{fig:nmmm}, top row.

The uncertainties of the model atmospheres are quite large.  For temperature,
the RMS fluctuations at ground level range between $3$~K during austral summer
to $6$~K during austral winter\footnote{Austral summer refers to the months of
December, January and February, austral winter corresponds to June, July and
August.}; at $26$~km, the RMS spread is $0.5$~K during austral autumn and
$5.0$~K during austral spring. Atmospheric depth varies mainly between $4$~km
and $8$~km.  The RMS spread of atmospheric depth at ground ranges between
$2$~\gcmsq (summer) and $5$~\gcmsq (winter); the largest RMS, at $8$~km, is
about $7.5$~\gcmsq.  Above $18$~km, the depth uncertainties are below
$1.5$~\gcmsq.  The vapor pressure RMS at ground is $1.5$~hPa (summer) and
$4.0$~hPa (winter), but is well below $0.2$~hPa above $7$~km. For illustration,
the uncertainties are plotted exemplarily for February (austral summer) and
August (austral winter) in Fig.~\ref{fig:nmmm}, bottom row.

\subsection{Optical Transmission and Cloud Detection}

During the 15 to 19 nights per lunar cycle
%numbers according to R. Smida's note GAP-2011-039
that are dark enough to operate the fluorescence telescopes, hourly measurements
of the aerosol optical depth~\cite{Abraham:2010,Valore:2009icrc} are made as a
function of altitude with two central laser facilities~\cite{Fick:2006yn} and
four lidar stations~\cite{BenZvi:2006xb}.  In addition, an optical telescope
called the ph(F)otometric Robotic Atmospheric Monitor (FRAM)~\cite{Prouza:2010}
is used to measure the integral aerosol optical depth inside and outside the
field of view of the FD building at Los Leones.  Finally, the cloud coverage at
the Observatory is measured with the lidar stations and infrared cameras located
at each of the four FD sites~\cite{Abraham:2010}.

There are four lidar stations, one per FD site, and during regular operations
the lidars are used to scan the atmosphere outside the field of view of the FD
telescopes. Currently, the scans are used to retrieve the mean cloud cover and
the lowest cloud height during each hour of FD measurements.  IR cloud cameras
provide complementary 2D images of the whole field of view every five
minutes~\cite{Abraham:2010}.  A direct combination of these two pieces of
information is used to provide a three-dimensional map of clouds above the
Observatory, but not without ambiguities. For instance, inspection of the lidar
data has shown that multiple cloud layers are present above the site about
$30\%$ of the time; a mismatched altitude may be associated to the clouds
detected by the IR cameras since different cloud layers cannot be easily
distinguished in the IR images.

FRAM is a robotic optical telescope with primary mirror diameter of 0.3~m
located about 30~m from the fluorescence detector building at Los Leones.  The
instrument was installed primarily to determine the wavelength dependence of the
extinction caused by Rayleigh and Mie scattering. This goal is achieved using
the photometric observations of selected standard (i\,.e.\ non-variable) stars,
and recently also using the photometric analysis of CCD images. The results of
this primary mission are presented in~\cite{BenZvi:2007icrc_fram}. Since its
installation in 2005, the FRAM telescope has also been involved in automatic
observations of optical transients of gamma-ray bursts.  This program is very
successful and several light curves of transients were already observed,
including one uniquely bright GRB afterglow~\cite{Jelinek:2006}.

\section{The Rapid Atmospheric Monitoring Program}
\label{sec:implementation}

Atmospheric uncertainties grow as a function of primary particle energy because
of the energy dependence of the longitudinal development of air showers, which
affects the geometry of observable showers within the field of view of the
FD~\cite{Abraham:2010}.  An improvement in the resolution of the atmospheric
monitoring data can be achieved by triggering measurements of the atmosphere
within a suitable time interval after the detection of high-energy showers above
a certain threshold (e.\,g., $E \gtrsim 10^{19}$~eV).  Such triggers have been
implemented for the individual weather balloon, lidar, and FRAM optical
telescope subsystems.

During the FD data taking, an automated process is used to collect event data from
the FD and SD, build and reconstruct hybrid events, and send the reconstructed
shower parameters to the atmospheric monitoring subsystems participating in the
rapid monitoring program.  Each subsystem performs individualized cuts on the
shower parameters, and if the shower is a candidate for special monitoring --
e.\,g., it has a well-observed track and is of a particularly high energy -- an
atmospheric measurement is performed either in the vicinity of the shower track
(for transmission measurements) or above the Observatory with meteorological
weather balloons.  In this manner, the time resolution of the atmospheric
measurements can be reduced from hours to minutes (for the lidars and the FRAM)
or from days to hours (for the weather balloons) with respect to the arrival
time of an interesting shower.  Moreover, the lidar stations and the FRAM are
able to directly probe the atmosphere along the shower-detector plane --~the
plane defined by the position of the FD telescope and axis of the shower~--
reducing the uncertainties introduced by the assumption of horizontally uniform
atmospheric layers in the weather databases.

The rapid atmospheric monitoring system consists of three components: an online
event builder that merges shower data as they are sent to the Observatory campus
in \mal; a hybrid reconstruction that uses all the detector and calibration data
that are available immediately after a shower is detected; and a broadcast
program that notifies the atmospheric subsystems of the detection of a hybrid
event.  The programs are designed to run without human intervention during FD
measurements.  We discuss the software components in
Section~\ref{subsec:onlineSoftware} and review the performance of the
reconstruction in Section~\ref{subsec:recoPerformance}.

\begin{figure}[tb]
  \begin{center}
    \includegraphics*[width=.95\textwidth,clip]{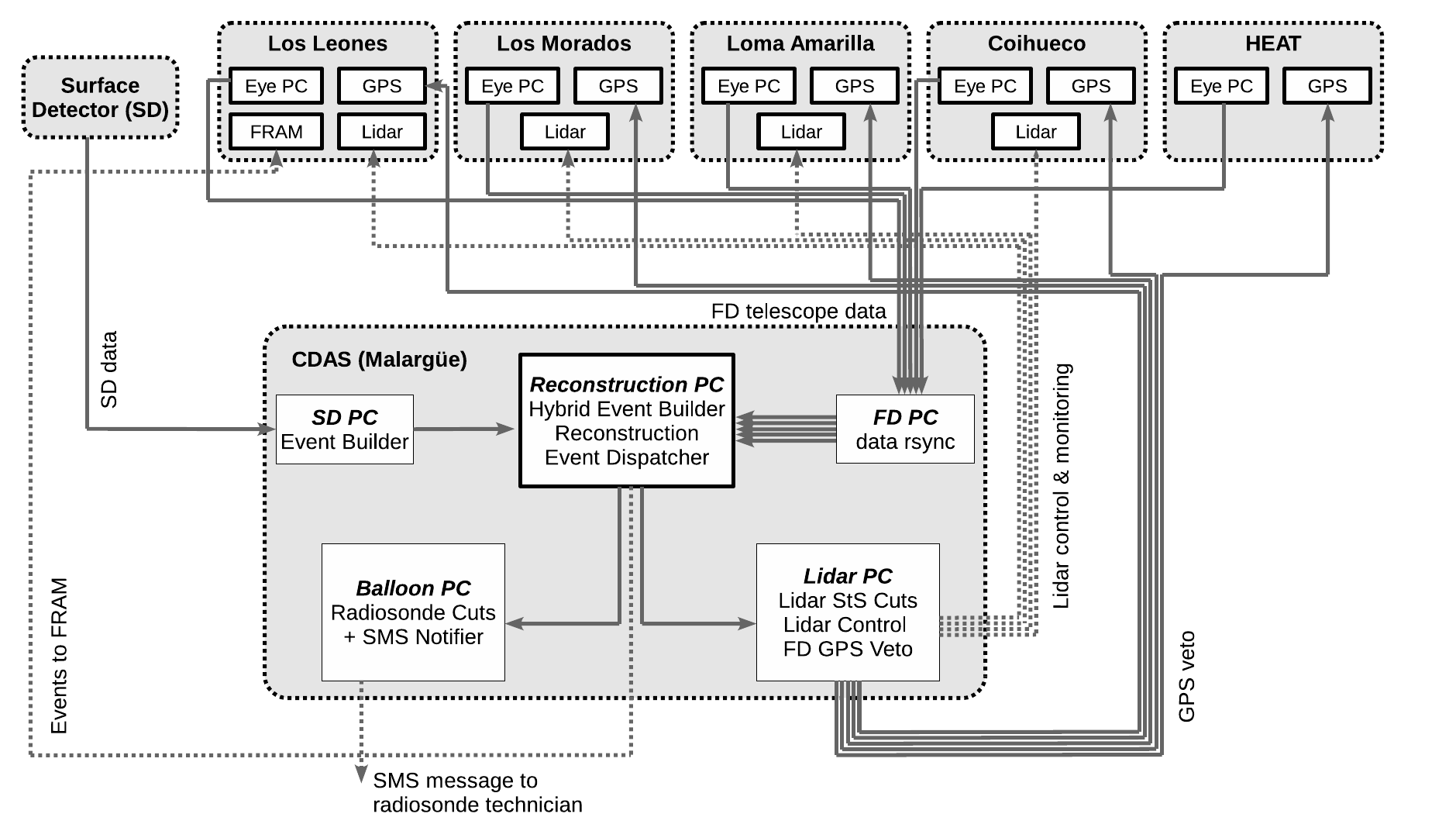}
    \caption{\label{fig:rapidMoniSystem} \slshape
      Network diagram of the rapid atmospheric monitoring system.  Data from the
      fluorescence and surface detectors are transferred to a Reconstruction PC
      in \mal.  The data are merged, reconstructed, and sent to the
      atmospheric monitoring PCs (Lidar PC, Balloon PC, and FRAM), where 
      triggers are formed and sent to the monitoring devices.  The
      Lidar PC also inhibits FD data acquisition during laser shots by sending
      a veto request to the FD GPS clock modules (see Section
      \protect \ref{subsec:lidarVeto}).}
  \end{center}
\end{figure}

\subsection{Online Event Builder, Reconstruction, and Broadcast
\label{subsec:onlineSoftware}}

The flow of data between the Observatory campus and the atmospheric subsystems
is shown in Fig.~\ref{fig:rapidMoniSystem}.  During FD measurement periods, data from the
fluorescence telescopes are transferred to \mal in a 20-second cycle.
Simultaneously, triggers and recorded data from the surface array are sent to
the SD Central Data Acquisition System (CDAS), a computer cluster and disk array
located in \mal.  Due to a polling delay that allows the SD communications
system to collect data from across the array, surface station data typically
arrive in the CDAS $2$ to $8$ minutes after the detection of an air shower.

Once the SD and FD data are available in \mal, a fast online event builder
produces air shower data in the standard hybrid format.  The data are
reconstructed using a version of the Auger reconstruction
software~\cite{Argiro:2007qg}, which is named \Offline, modified for online
running. The online reconstruction is configured to use the latest available
detector and calibration databases, and it is kept in sync with releases of the
\Offline software.  This is to keep the results of the online reconstruction as
close as possible to the standard offline\footnote{While \Offline will refer
only to the software framework, ``offline'' is meant to describe processes that
happen several days to months after the measurement.} reconstruction.  However,
since the non-event databases are typically updated on timescales of 4-6 months,
some drifts between the online and offline reconstructions are unavoidable.

In the offline reconstruction, large-particle scattering by aerosols is
estimated using atmospheric measurements. It is not possible to use real-time
atmospheric monitoring data in the online reconstruction, so instead an average
parametric model of aerosol scattering in \mal is used. Rayleigh scattering by
molecules is calculated using the \nmmm average monthly models. The systematic
uncertainties introduced by the use of average models is discussed in
Section~\ref{subsec:recoPerformance}.

Approximately 80 geometry, quality, and energy parameters from each
reconstructed shower are written to disk on the \emph{Reconstruction PC}
(cf.\ Fig.~\ref{fig:rapidMoniSystem}).  As they are saved to disk, the events
are also transferred to the atmospheric monitoring subsystems (\emph{Balloon
PC}, \emph{Lidar PC}, and \emph{FRAM}) via network broadcast.  Client programs
in the atmospheric monitors are used to perform cuts on the hybrid data and
issue triggers based on the specialized measurements performed with each
instrument (see Sections \ref{sec:bts}, \ref{sec:sts}, and \ref{sec:fram}).

\begin{figure}[tb]
  \begin{center}
    \includegraphics*[width=.95\textwidth,clip]{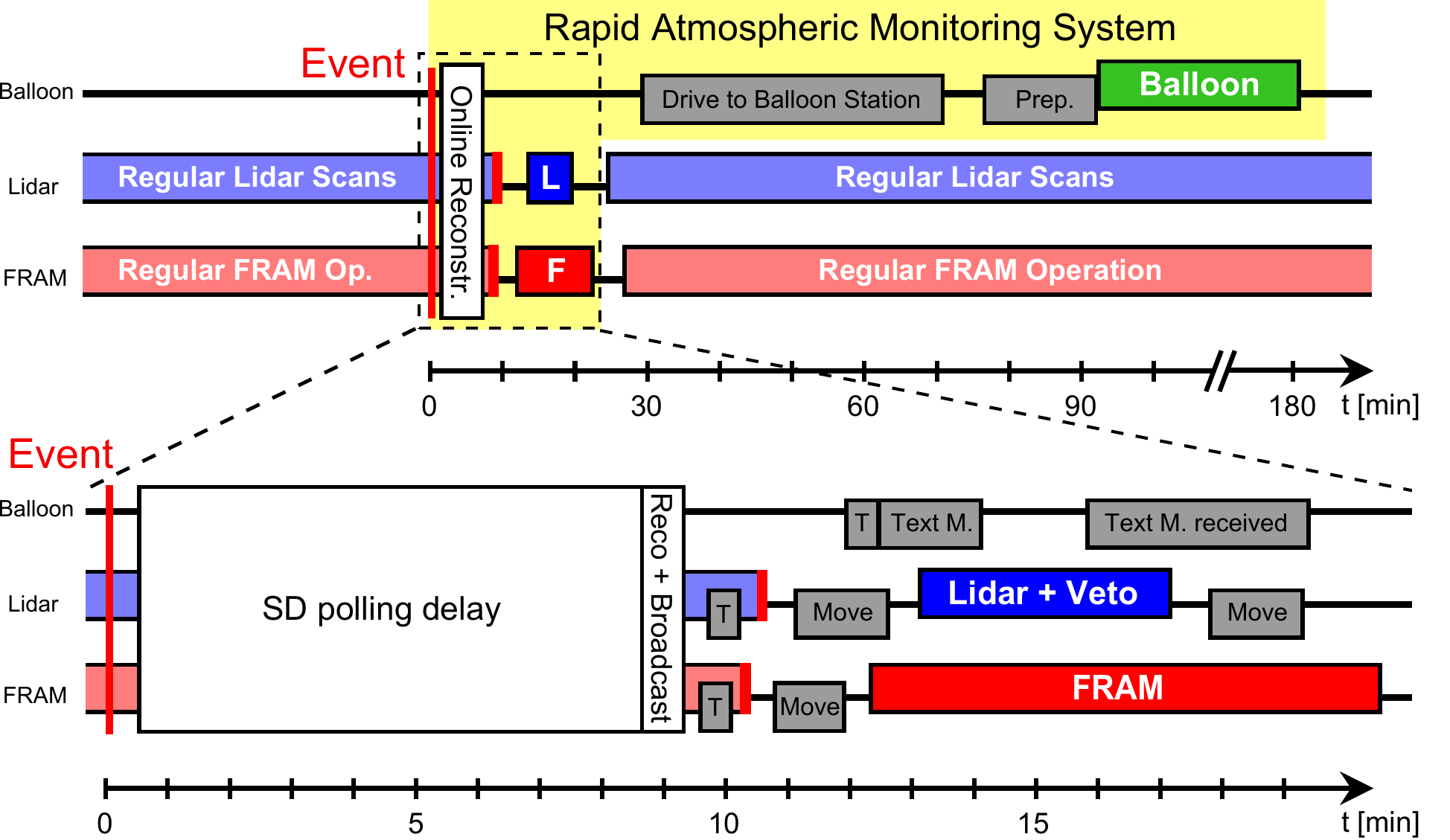}
    \caption[]{\label{fig:timing} \slshape
      The timing scheme of the rapid monitoring system.  \textup{Top:} the
      timeline of rapid monitoring with the lidar stations (L), FRAM (F), and
      the balloon launches in the context of standard operations.  Atmospheric
      scanning with the lidars and FRAM is typically completed within 20
      minutes of the detection of an event.  The balloon launches are initiated
      about 90 minutes after the event.  \textup{Bottom:} a detailed view of
      the first 20 minutes after the detection of an event.  Events are
      reconstructed within 2 to 9 minutes of the initial detection and the
      results are broadcast to the monitoring PCs.  Each subsystem applies an
      individualized trigger criterion (T) to identify showers for follow-up
      monitoring.  In the case of the lidar and FRAM telescopes, the regular
      operations are interrupted and the telescopes slew into position to begin
      a scan of the shower-detector plane.  During the lidar scan the FD DAQ is
      inhibited by a veto to avoid spurious triggers caused by scattered laser
      light.  Regular atmospheric sweeps resume once the scans are complete.
      In the case of the balloon system, a text message is sent to an on-site
      technician who drives to the balloon launch facility, prepares the
      balloon, and starts the radiosonde measurement.  For more details, see
      Sections \ref{sec:bts}, \ref{sec:sts}, and \ref{sec:fram}.
    }
  \end{center}
\end{figure}

In Fig.~\ref{fig:timing}, a timing diagram is shown for the online
reconstruction and the activity within all three subsystems. More details on the
individual steps are provided in the corresponding sections. It should be noted
that the online reconstruction runs continuously. The pictured timeline shows
only the case if an interesting air shower event is identified by subsequent
steps. Also, the three systems operate independently, they do not necessarily
trigger on the same air shower event because of different trigger criteria.

\subsection{Reconstruction Performance
\label{subsec:recoPerformance}}

We illustrate the performance of the online reconstruction using hybrid data
recorded between March 2009 and March 2011.  During this period, 320 hybrid
events reconstructed online had energies above $10^{19}$~eV and passed standard
quality cuts based on the event geometry and d$E$/d$X$ profile
fit~\cite{Abraham:2010yv, Abraham:2010mj}.  Applying the same cuts to data
reconstructed offline produces a set of 382 events during the same period.

Inspection of the events shows that the online and offline sets have only 255
events in common.  The discrepancy, and the lower number of online events, is
caused by several factors.  The number of events reconstructed online is reduced
by downtime in the online reconstruction due to various technical problems such
as software failures, network crashes, etc.  For example, the downtime of the
online reconstruction during 2010 was about 15\%, which accounts for much of the
difference in size between the online and offline event samples.  In addition,
most of the offline data were corrected for real aerosol conditions, whereas the
online reconstruction uses an average model of aerosols above the Observatory.
The shower profiles reconstructed online tend to be of worse quality because
true aerosol scattering is not taken into account, and so more events fail the
offline quality cuts on the shower profile.  The migration of events around the quality
cuts due to changes in the software versions and databases used in the
reconstruction also accounts for an additional reduction in the number of events
in common between the online and offline data sets.

\begin{figure}[tb]
  \begin{center}
    \includegraphics*[width=.93\textwidth,clip]{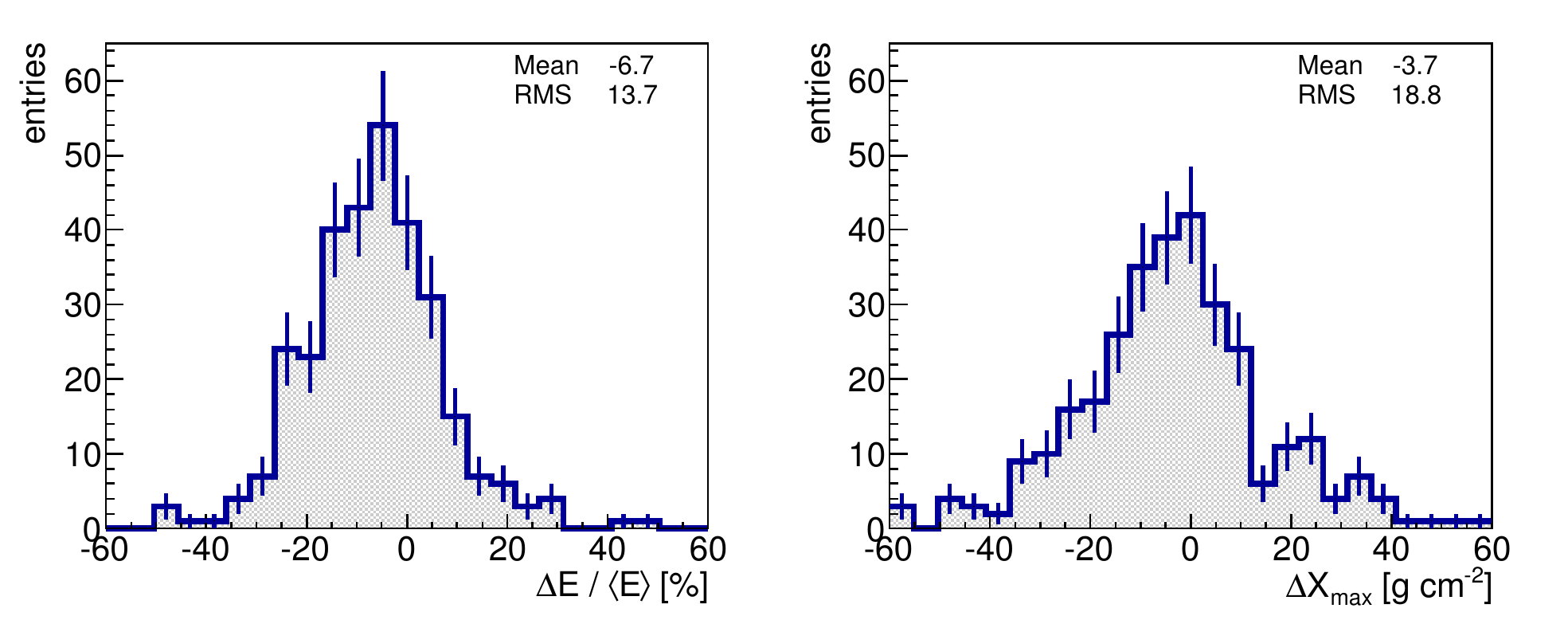}
    \caption{\label{fig:onlineVsObserver} \slshape
      Comparison of shower energy and position of shower maximum reconstructed
      online and offline.  Note that $\Delta E=E_\text{offline}-E_\text{online}$
      and $\Delta
      X_\text{max}=X_\text{max}^\text{offline}-X_\text{max}^\text{online}$.
      $\langle E\rangle$ is the average of the two reconstructed energies.
      Real hourly aerosol measurements were used in the offline
    }
  \end{center}
\end{figure}

It is instructive to compare the common events of the two data sets.  In
Fig.~\ref{fig:onlineVsObserver}
the differences in the energy and \xmax of the common events are plotted.  Both
distributions contain significant tails, and the energy reconstructed online is
systematically higher than the energy reconstructed offline by about 7\%.  The
main cause is the lack of true aerosol corrections in the online data, which
accounts for at least half the offset between the two
reconstructions~\cite{Abraham:2010}.
The remainder of the offset is due to differences in software versions
between the online and offline reconstructions and the lack of nightly
calibration constants in the online reconstruction.

Even though the online reconstruction is affected by a non-negligible downtime,
it appears to have performed well since it was first implemented in 2009.  The
comparison between the online and offline events indicates the presence of a
significant systematic bias in the online data because of the use of an average
aerosol model.  This means that some events which pass the
online cuts may not survive the offline analysis cuts. In the
absence of real-time aerosol data this is unavoidable. However, it may be
possible to tune certain measurements using nearly real-time conditions and
hence reduce ``false positive'' triggers.  An example application is discussed
in Section~\ref{subsec:dbtrigs}.

\section{{\textit{\textbf Balloon-the-Shower}} Program
\label{sec:bts}}

The use of monthly site models to estimate atmospheric state variables and
molecular scattering rather than real-time radiosonde data introduces an
uncertainty into the estimated production and transmission of fluorescence
light in air showers. This contributes to the statistical uncertainties in the
reconstructed energy and position of shower maximum.  The total effect is
moderate, but it does depend on the shower energy.  Between primary energies of
$10^{17.7}$~eV and $10^{20}$~eV, the monthly profiles contribute $1.5\%$ (at
$10^{17.7}$~eV) and $3\%$ (at $10^{20}$~eV) to the total energy resolution of
about $8\%$~\cite{Dawson:2007}, and \mbox{$7.2$--$8.4$~\gcmsq} to the total 
\xmax resolution of about 20~\gcmsq~\cite{Abraham:2010yv} of the hybrid
reconstruction~\cite{Abraham:2010,Keilhauer:2009icrc2,Keilhauer:2009icrc}. It is
important to note that these numbers are characteristic of a large sample of
showers, but the systematic errors in the reconstruction of individual showers
can be substantially larger, particularly at high energies.  Therefore, it is
desirable to minimize as much as possible the atmospheric uncertainties in the
reconstruction of high-energy events.

To improve the resolution of the reconstruction for the highest-energy showers,
the \emph{Balloon-the-Shower} program (\bts) was operated between March 2009 and
December 2010.  Its purpose was to perform an atmospheric sounding within about
three hours of the detection of a high-quality high-energy event.

\subsection{Performance of \bts
\label{sec:bts_perform}}

In March 2009, \bts replaced regularly scheduled meteorological radio soundings
at the Observatory.  The target launch rate was chosen to be three to seven
launches per FD measurement period, with each period lasting about 2.5 weeks.
The focus of the \bts program was high-energy showers used in the SD energy
calibration or the hybrid mass composition analysis; in other words, hybrid
events with well-reconstructed longitudinal profiles and energies above
$10^{19}$~eV.

\begin{figure}[tbp]
  \centering
  \includegraphics[width=1.\textwidth,clip]{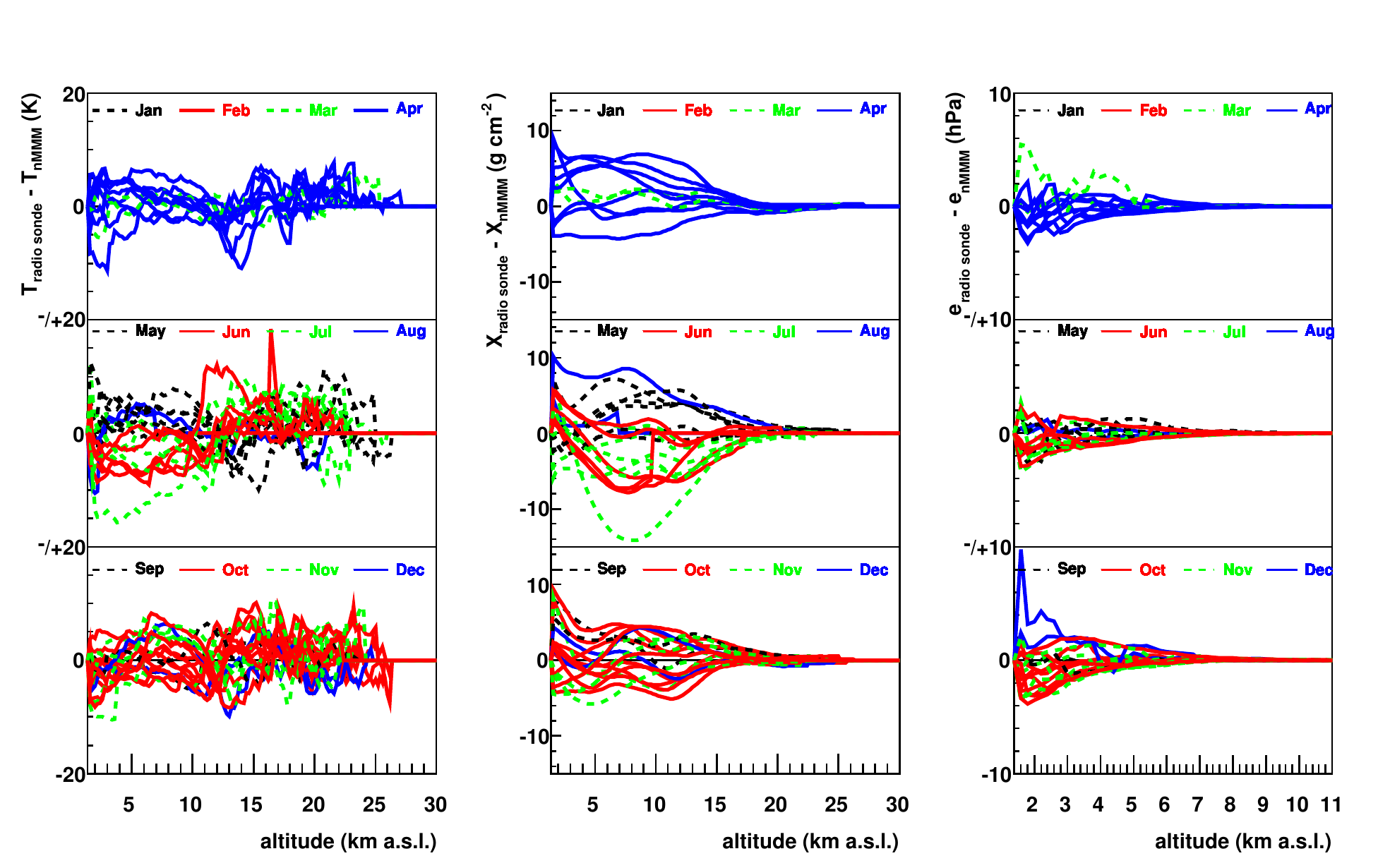}
  \caption{\label{fig:atmos_bts} \slshape
    Atmospheric profiles obtained from launches within the \bts program for temperature
    (left), atmospheric depth (middle), and water vapor pressure (right). Each
    actual profile is shown in difference to the according profile
    from the \nmmm. There are no launches for January and February.}
\end{figure}

The atmospheric profiles from the \bts program represent an independent data
set that can be compared to the \nmmm average models. The difference between
each \bts profile and its corresponding \nmmm profile is plotted in
Fig.~\ref{fig:atmos_bts}. The width of the deviations is in agreement with the
uncertainties of the monthly models described in Section~\ref{subsec:nMMM}.

Events passing the online cuts were used to trigger a text message sent to an
on-site technician, who then drove to the Balloon Launch Station to launch a
weather balloon.  Given the lack of automation, the radiosonde flights typically
took place only several hours after the detection of a cosmic ray event.  To
minimize the delay, it was decided to limit the time difference
between event detection and balloon launch to a maximum of three hours.  This
delay was not expected to affect the validity of the radiosonde data, since
fluctuations in the vertical atmospheric profiles tend to be much larger between
nights than within a single night~\cite{Wilczynska:2006wt}.

\subsubsection{Quality Cuts
\label{sec:bts_cuts}}

To trigger a \bts launch, showers from the online reconstruction were required
to pass quality cuts used in publications of the SD energy
spectrum~\cite{Abraham:2010mj} and the hybrid mass
composition~\cite{Abraham:2010yv}.  The cuts are listed in Table~\ref{t:btsCuts}
and were designed to minimize the uncertainty in shower energy and \xmax.  In
fact, the cuts used for \bts are moderately stricter than those used
in~\cite{Abraham:2010yv,Abraham:2010mj} to account for the systematic
uncertainties in the online reconstruction described in
Section~\ref{subsec:recoPerformance}.

Cuts~(1) and (2) select showers in which the reconstructed energy and the
position of shower maximum are reliably estimated.  Cut~(3) removes showers in
which \xmax is less than $20$~\gcmsq from either the minimum or maximum observed
depth of the shower track.  This reduces the possibility that \xmax is
mis-identified and also improves the reconstructed d$E$/d$X$ profile.  Cut~(4)
is a standard geometry cut that ensures the surface station with the largest
signal (i.\,e., the one used for timing in the hybrid reconstruction) is close
to the shower core.  Cuts~(5) and (6) are indicators of the quality of a
Gaisser-Hillas parametric fit to the longitudinal shower
profile~\cite{Unger:2008uq,Gaisser:1977}.  Cut~(5) is effective at removing
showers obscured by clouds and other atmospheric non-uniformities, since the
profiles of such showers deviate strongly from a Gaisser-Hillas curve. This cut
also removes possible exotic air shower candidates that are covered by the other
two subsystems (cf. Sec.~\ref{subsec:dbtrig} and Sec.~\ref{subsec:dbtrigfram}),
as they are not the main focus of the \bts program.  Cut~(6), a $\chi^2$
difference between a linear fit and a Gaisser-Hillas fit to the longitudinal
profile, removes faint, low-energy showers from the trigger sample.  The
fraction of rejected events of the cuts are included in Table~\ref{t:btsCuts}.

\begin{table}[tb]
  \begin{center}
    \begin{tabular}{l l c}
      \toprule
      Quality Cut                 &                                           & Rejected Events \\
      \midrule
      (1) Energy uncertainty      &  $\sigma_\mr{E}/E~<~0.2$                           & 47.5\% \\
      (2) \xmax uncertainty       &  $\sigma_{X_\text{max}}~<~40~\text{\gcmsq}$        & 59.5\% \\
      (3) Field of view           &  $X_\mr{max}$  well observed                       &  8.8\% \\
      (4) Distance to SD with highest signal & $d_\mr{axis}~<~1500~\text{m}$           &  0.5\% \\
      (5) Quality of Gaisser-Hillas (GH) fit & $\chi^2_\mr{GH} / n_\mr{dof}~<~2.5$     &  5.3\% \\
      (6) Comp.\ of GH with linear fit       & $ \chi^2_\mr{lin} - \chi^2_\mr{GH}~>~4$ & 36.8\% \\
      \midrule
      Energy threshold            &  $E_0 > 19.95$~\EeV                                & 99.3\% \\
      \bottomrule
    \end{tabular}
    \caption{\label{t:btsCuts}
      A list of quality cuts for the \bts program.  The fraction of rejected events were
      calculated using hybrid data recorded from 2006 to 2009.  Note that the
      percentage in each row is given with respect to the previous cut.
      In the last line, all events below the energy threshold were discarded.
    }
  \end{center}
\end{table}

%Efficiency = part of the selected useful events from useful events
%Purity = percentage of the selected useful events to the total selected events
%
%Year     All     E    Axis   Xmax   Xerr   Eerr    Chi2    Lin  Sel
%2009  143189  1232  140472  72378  38058  83454  134728  37245   86
%2008  131451   744  128342  63780  30300  69854  118526  32293   69
%2007  125140   890  123533  60632  29208  66065  113208  32213   64
%2006  100545   621   98830  48761  23708  53916   91858  26133   49
%
%2010  128026   724  126215  65200  33894  74861  121559  34025   84

\begin{figure}[htb]
  \begin{center}
    \includegraphics*[width=.59\linewidth,clip]{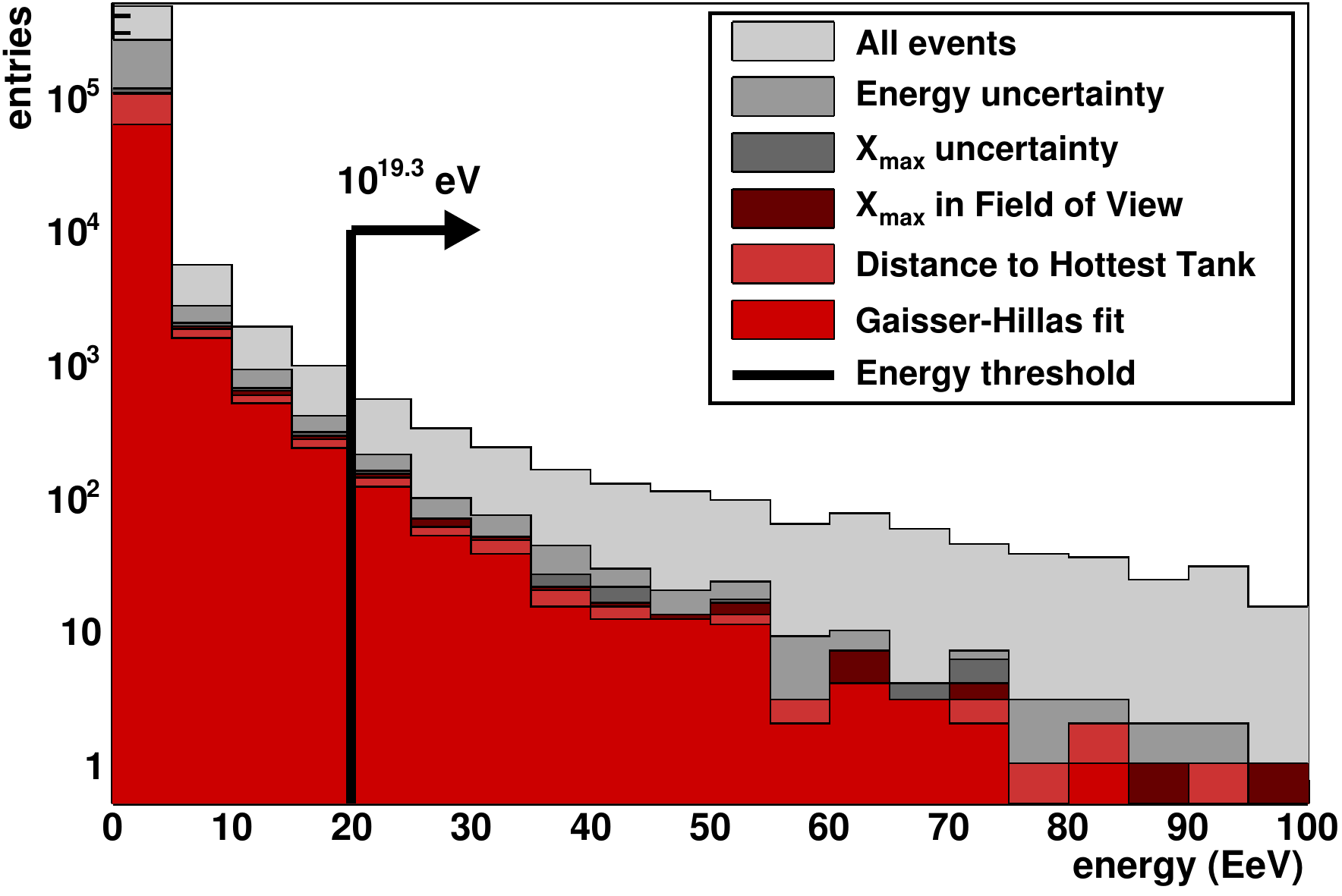}
    \caption{\label{fig:btsCuts} \slshape
      Candidate event distribution for \bts estimated by using hybrid events
      collected between 2006 and 2009. The effect of the quality cuts and the
      energy threshold is shown.  Note that the two profile fit cuts (5)
      and (6) were combined.
    }
  \end{center}
\end{figure}

The hybrid events recorded between January 2006 and January 2009 were used as a
tuning sample to set the rate of \bts triggers.  The effect of
the cuts are shown as a function of energy in Fig.~\ref{fig:btsCuts}.  To reach
the desired number of atmospheric soundings --~$50$ to $60$ per year, or about
$3$ to $7$ per FD measurement period~-- it was necessary to further reduce the
size of the event sample with a cut on the reconstructed shower energy.  The
energy threshold, determined from the tuning sample, was
\begin{equation}
  \log_{10}\left(\frac{E_\mr{min}}{\text{eV}}\right) =
  19.3~~~\Leftrightarrow~~~E_\mr{min} = 19.95~\EeV.
\end{equation}

\subsubsection{Trigger of Weather Balloon Launches
\label{sec:sms_trigger}}

After a shower passed the automatic \bts trigger, a short message containing
the date and time of the event was sent to the mobile phone of an on-site
technician.  If the message was received, the technician would drive to the
Balloon Launch Station and proceed with the atmospheric sounding within three
hours of the event.

Between March 2009 and December 2010, 100~text messages were sent to the
technician.  From these 100 triggers, 52~balloons were launched successfully.
Some messages were received while a radiosonde was already in flight, due to the
tendency of high-quality, high-energy observations to cluster during very clear,
cloudless nights.  Therefore, 62~\bts triggers were covered by the 52~flights.
The remaining triggers, about one-third of the total, were lost due to technical
issues such as a hardware failure at the Balloon Launch Station in August 2009
(11~events), problems with the transmission of the text messages,
or other failures in the radiosonde flights.

\begin{figure}[tb]
  \begin{center}
    \includegraphics*[width=.49\linewidth,clip]{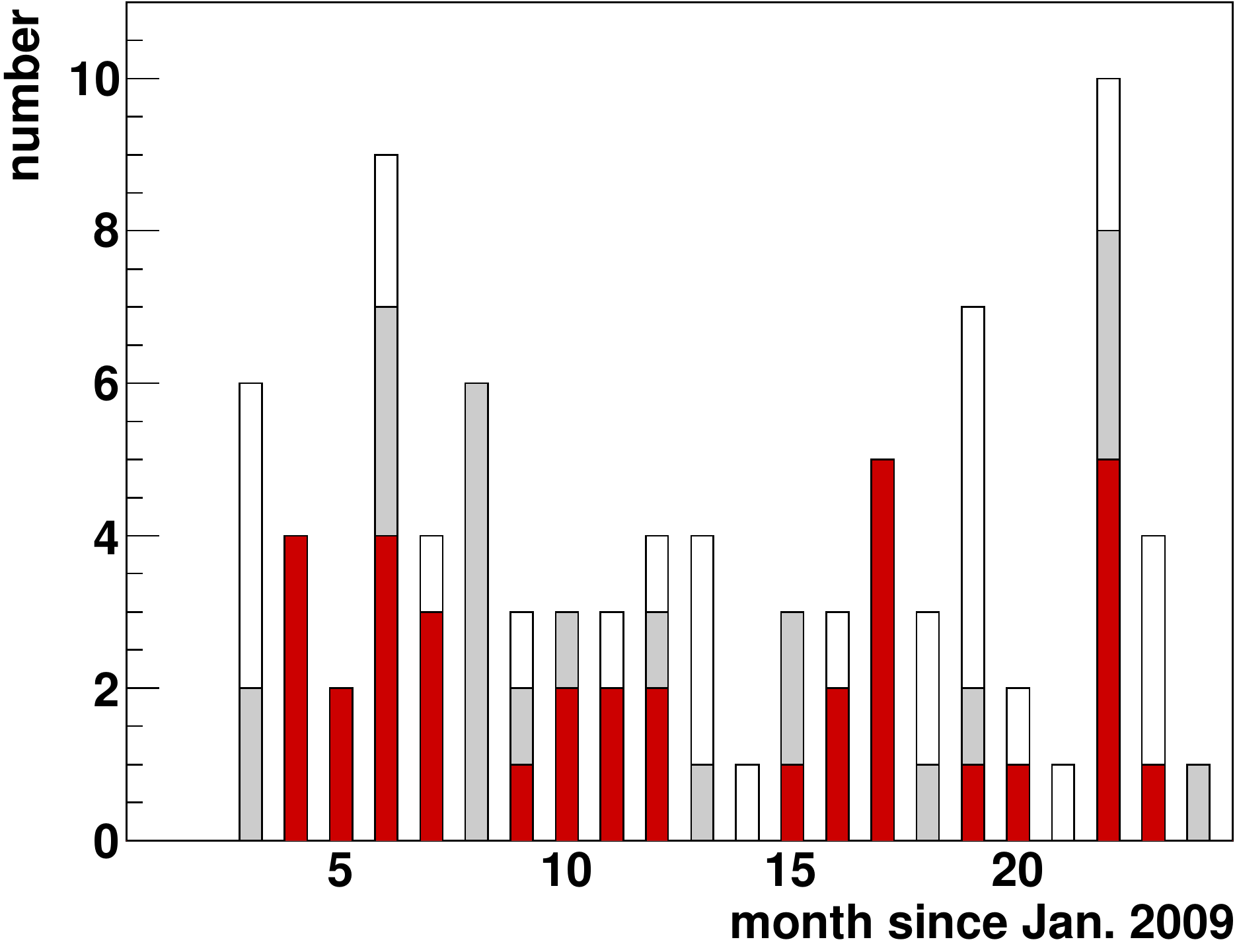}
    \caption[]{\label{fig:btsTriggers} \slshape
      Expected \bts triggers for the 22 measurement periods between March 2009
      and December 2010 using the offline reconstruction~\cite{Argiro:2007qg}
      and \bts cuts (Table~\ref{t:btsCuts}).  Also shown in grey is the number
      of generated text messages for those expected events. Shown in red
      is the number of events that were covered by a balloon launch. No visible
      white bar means every interesting event was caught by the system. No
      visible grey bar means there was a launch for every text message.
    }
  \end{center}
\end{figure}

The \bts statistics between March 2009 and December 2010 are shown in
Fig.~\ref{fig:btsTriggers}. Note that the chart shows events reconstructed
using the offline reconstruction (and after quality cuts were applied) and not
the online reconstruction.  During the period of \bts operations, 88~events
reconstructed offline passed the \bts cuts.  Of these, 59 were also identified
online, meaning that out of the 100~events which triggered a text message, 41 do
not survive the \bts cuts after offline reconstruction.  Some did not satisfy
the \bts cuts, while others fell below $10^{19.3}$~eV. Of the 59 common events,
35 were covered by a weather balloon launch.  If the energy cut is relaxed
slightly to $10^{19.2}$~eV, 51~offline events are covered by a balloon launch.

\subsection{Air Shower Reconstruction using \bts Data
\label{sec:bts_results}}

To evaluate the effectiveness of the \bts program, we have reconstructed hybrid
events covered by the radiosonde launches conducted since March 2009.  The
results are compared to a reconstruction that uses the \nmmm average monthly
conditions, as well as more real-time conditions given by the Global Data
Assimilation System (GDAS)~\cite{GDAS:2004}.  GDAS is a global atmospheric
model based on meteorological data and numerical weather predictions.
Altitude-dependent profiles of atmospheric state variables such as $T$, $p$,
and $e$ are provided on a $1^\circ\times1^\circ$ latitude-longitude grid.  The
GDAS database contains profiles which are useful for the needs of the Pierre
Auger Observatory with a time resolution of three hours starting in June 2005.
Hence, the database covers not only the period of \bts launches, but also most
of the period of data-taking at the Auger Observatory.  GDAS data were made
available to the air shower analysis of the Auger Observatory beginning in
spring 2011~\cite{Keilhauer:2011}.

\subsubsection{Effect of \bts Profiles and Model Atmospheres on the
Reconstruction
\label{subsec:BtSvsMMM}}

To study the effect of the \bts data on the reconstruction of air shower
profiles, we have reconstructed the 62~hybrid events covered by 52~\bts
launches.  This data sample contains 52 events which
pass all quality cuts consisting of 90 individual fluorescence profiles after 
accounting for events observed in stereo with
multiple telescopes.  We also compared the reconstruction using \bts data to
those using \nmmm and GDAS model profiles. In case of reconstructions using
\nmmm, two more events where discarded by cut criteria, thus only 50 events are
contained in this reconstruction data sample. In all cases, we have accounted not
only for the effects of the atmospheric profiles on light scattering, but also
for the effects of temperature, pressure, and humidity on fluorescence light
production~\cite{Arqueros:2008,Abraham:2010,Keilhauer:2010ecrs}.  For the
fluorescence light calculation, experimental data from the AIRFLY
experiment~\cite{Ave:2008} and conference contributions~\cite{Bohacova:2009afw}
from the AIRFLY collaboration were used.

\begin{figure}[htbp]
  \centering
  \includegraphics[width=.48\textwidth,clip]{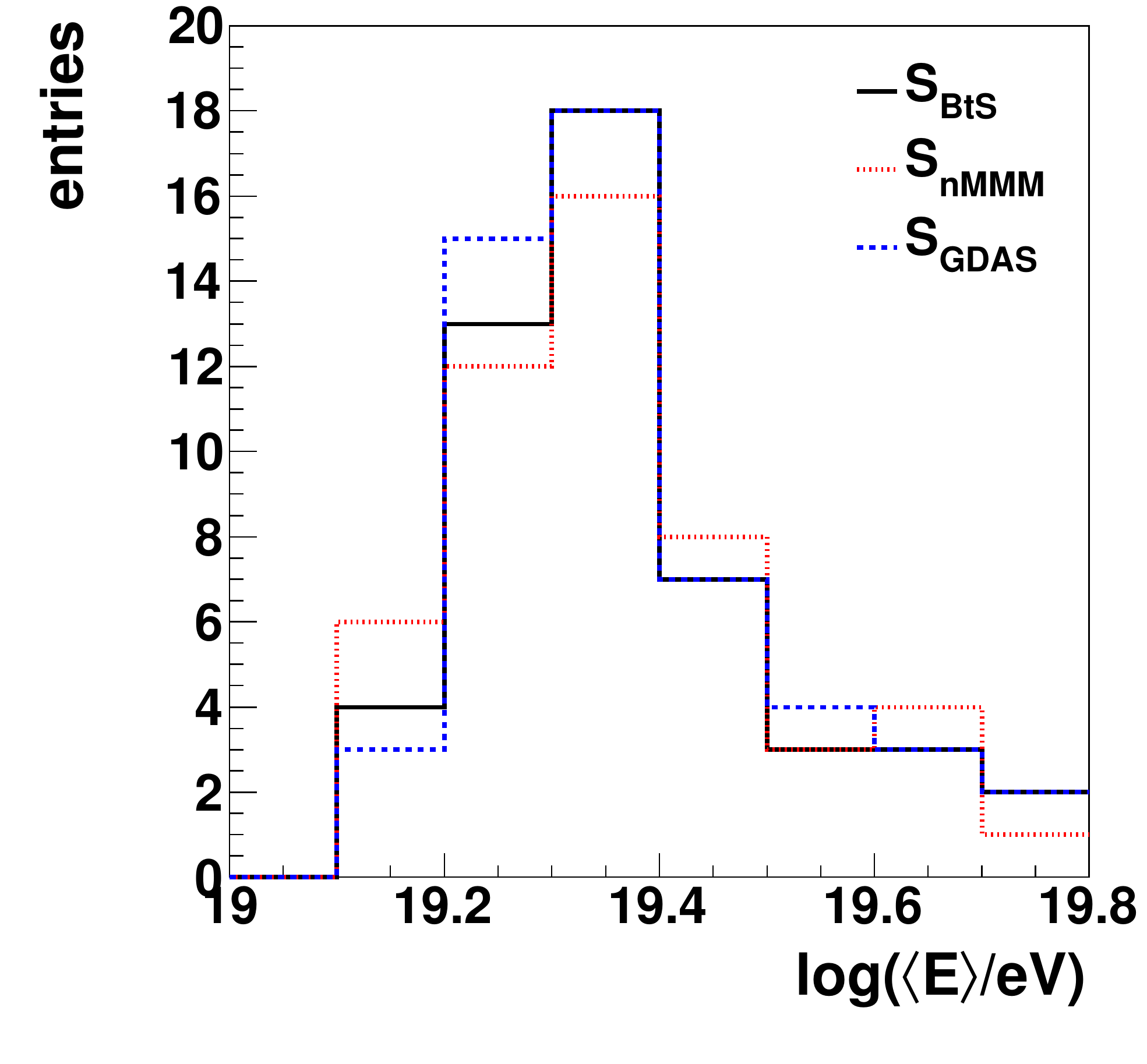}
  \caption{\label{fig:energydist} \slshape
    Energy distribution of all reconstructed air showers
    which passed the \bts quality cuts during the online analysis and for
    which a radio sounding has been performed.  The black, solid line represents
    the energy distribution of 52 events reconstructed offline with actual
    atmospheric conditions as measured during the dedicated weather balloon
    ascents, \sbts. The red, dotted line, \snmmm, displays 50 successfully
    reconstructed events and the blue, dashed line --~\sgdas~-- indicates
    the distribution of the same 52 events as in \sbts. The two missing
    events not covered in the offline reconstruction \snmmm are due to onset
    of a cut criterion or due to failed reconstructions because of extreme shower
    geometries. The distributions agree very well within the expected uncertainties.}
\end{figure}

In the study, we devote our main attention to the reconstructed cosmic ray
energy $E$ and depth of shower maximum \xmax of the air shower profiles.  The
energy distribution of the reconstructed events is provided in
Fig.~\ref{fig:energydist}. For air showers detected by more than one telescope,
the weighted mean of the shower observables is used. All events represented by
the solid line are reconstructed using the atmospheric profiles gathered within
the \bts program, \sbts. Reconstructions applying the monthly atmospheric
conditions as described by \nmmm, \snmmm, are displayed with a dotted red line.
The third set of reconstructed air showers, \sgdas, is plotted as a dashed blue
line and was obtained using the corresponding model atmospheres from GDAS.  The
three distributions agree well within the systematic uncertainty of the hybrid
energy reconstruction~\cite{Dawson:2011zz}. The overall systematic uncertainty
is 22\%, whereof 1\% are contributions due to atmospheric
uncertainties~\cite{Pesce:2011icrc} which are discussed here. Note that some
events have spilled below the energy threshold of $10^{19.3}$~eV because of the
systematic energy shift between the online and offline reconstructions.  The
mean energy of the event sample is $10^{19.4}$~eV.

\begin{figure}[tb]
  \hfill
  \begin{minipage}[t]{.48\textwidth}
    \centering
    \includegraphics*[width=\linewidth]{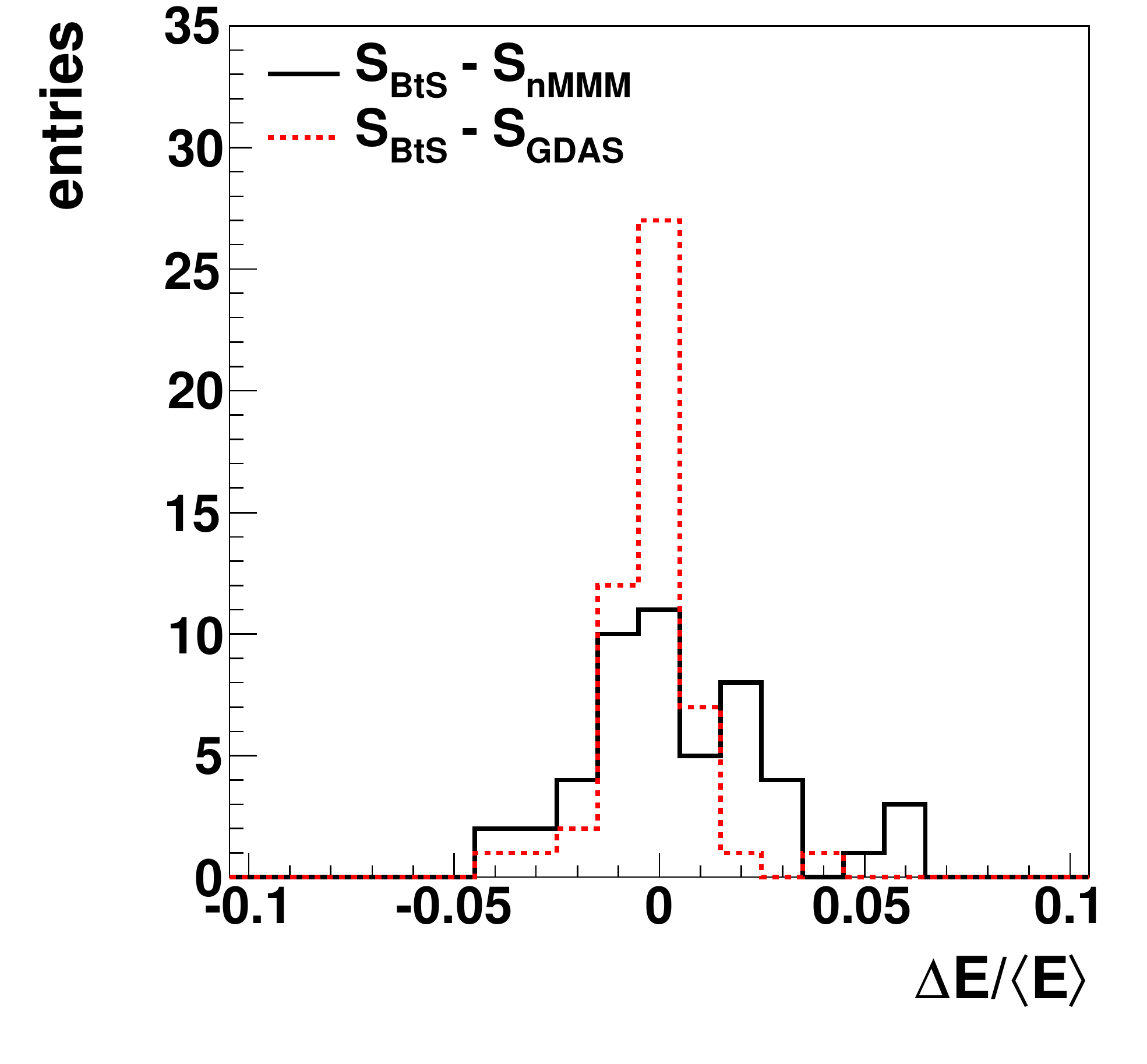}
    \caption{\label{fig:dE} \slshape
      Energy difference between events selected by the \bts program. The air
      showers are reconstructed with three different atmospheric descriptions,
      for details see text.
    }
  \end{minipage}
  \hfill
  \begin{minipage}[t]{.48\textwidth}
    \centering
    \includegraphics*[width=\linewidth]{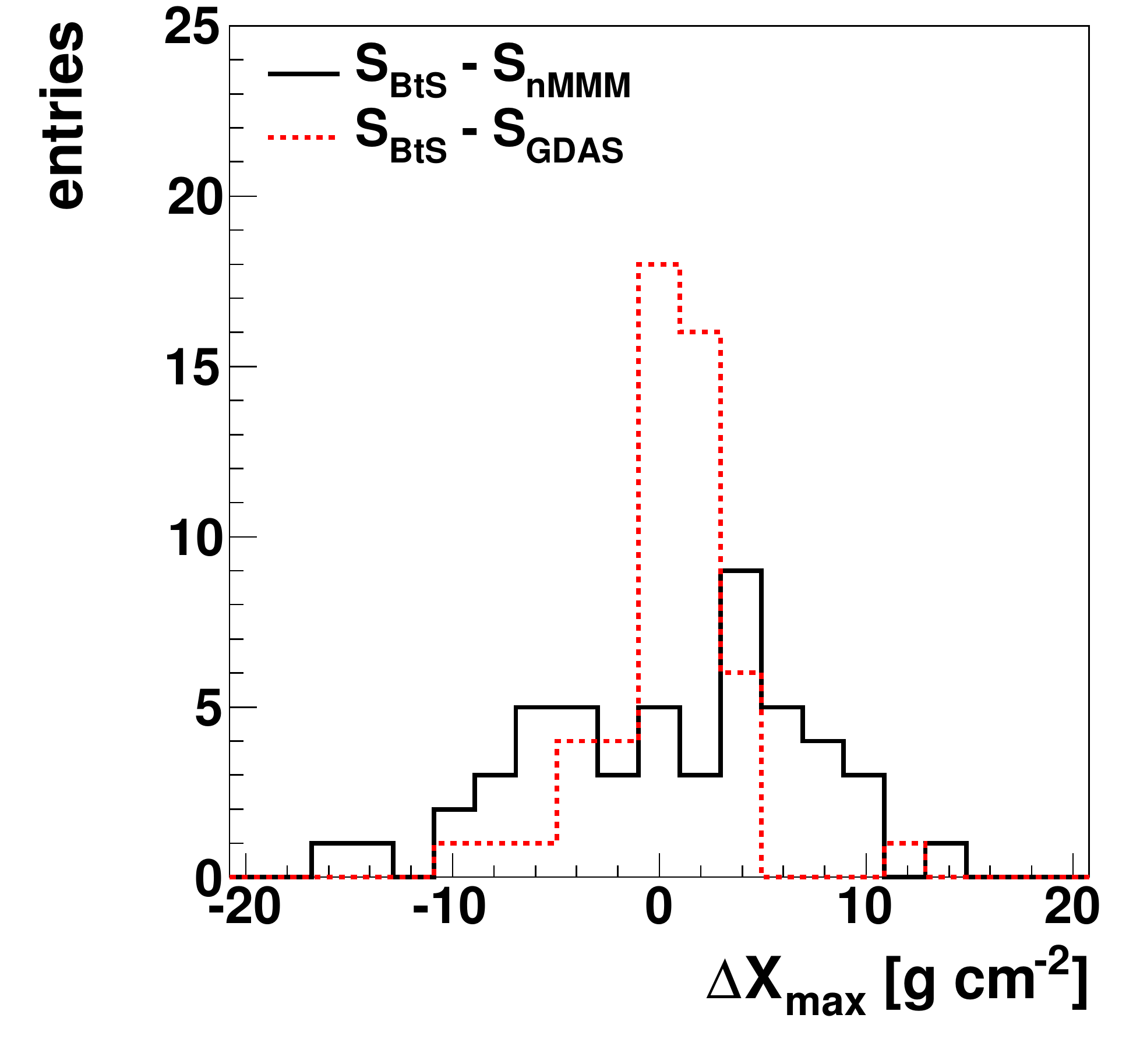}
    \caption{\label{fig:dXmax} \slshape
      Difference of the position of shower maximum \xmax between events selected
      by the \bts program. The air showers are reconstructed with three
      different atmospheric descriptions, for details see text.
    }
  \end{minipage}
\end{figure}

\begin{table}[tb]
  \begin{center}
    \begin{tabular}{lcccc}
      \toprule
      & \relEavg [\%] & RMS(\relEavg) & \relxmax [\gcmsq] & RMS(\relxmax) [\gcmsq] \\
      \midrule
      BtS -- nMMM & ~~~0.5 & 2.3 & 0.3 & 6.6 \\
      BtS -- GDAS & $-$0.2 & 1.2 & 0.5 & 3.3 \\
      \bottomrule
    \end{tabular}
    \caption{\label{t:recomeanrms}
      The mean differences and RMS values for the comparisons of \sbts and
      \snmmm as well as \sbts and \sgdas in relative energy (\relEavg) and
      position of shower maximum (\relxmax).
    }
  \end{center}
\end{table}

In Fig.~\ref{fig:dE} and Fig.~\ref{fig:dXmax}, the distributions of the energy
difference \relEavg and \relxmax between the \sbts and \snmmm events (solid
line) and the \sbts and \sgdas events (dotted line) are plotted, respectively.
The quantity $\langle E\rangle$ is the average of the energies reconstructed
with each pair of atmospheric profiles.  The mean differences and widths of the
distributions for both \relEavg and \relxmax are listed in
Table~\ref{t:recomeanrms}.
For the \bts-\nmmm comparison, the most extreme differences of about $6\%$ and
$16$~\gcmsq are found for \relEavg and \relxmax, respectively.  The width of the
distribution for the \bts-GDAS comparison is smaller because the time resolution
of the GDAS profiles is much finer than that of the monthly models.  The
comparison indicates that the GDAS data provide a reasonable description of the
local conditions on time scales of a few hours.

Finally, it can be concluded that the systematic uncertainties of the energy
(22\%) and of \xmax (13~\gcmsq~\cite{Abraham:2010yv}) are hardly reduced
by applying actual atmospheric profiles in the reconstruction of extensive 
air showers instead of applying adequate local models. The resolution of $E$
and \xmax can be slightly reduced by 0.3\% and 1.1~\gcmsq, respectively.

%In Fig.~\ref{fig:dE}, the distributions of the energy difference \relEavg
%between the \sbts and \snmmm events (solid line) and the \sbts and \sgdas
%events (dotted line) are plotted. The quantity $\langle E\rangle$ is the
%average of the energies reconstructed with each pair of atmospheric profiles.
%For the \bts-\nmmm comparison, the mean difference in the reconstructed energy
%is $0.5\%$ and the width of the distribution is $2.3\%$, with the most extreme
%difference of about $6\%$.  For the \bts-GDAS comparison, the mean difference
%in the reconstructed energy is $-0.2\%$ with an RMS of $1.2\%$.  The width of
%the distribution is smaller because the time resolution of the GDAS profiles is
%much finer than that of the monthly models.  The comparison indicates that the
%GDAS data provide a reasonable description of the local conditions on time
%scales of a few hours.
%
%The distributions of the difference \relxmax between the reconstructions are
%shown in Fig.~\ref{fig:dXmax}, where the \sbts-\snmmm comparison is plotted as
%a solid line and the \sbts-\sgdas comparison is drawn with a dotted line.  The
%mean of the \sbts-\snmmm distribution is $0.3$~\gcmsq with an RMS of
%$6.6$~\gcmsq, with individual differences as large as $16$~\gcmsq.  The \sbts
%and \sgdas reconstructions appear to be more compatible, with a mean
%of $0.5$~\gcmsq and a width of $3.3$~\gcmsq.

\subsubsection{Study of Systematics
\label{sec:bts_syst}}

\begin{figure}[htbp]
  \begin{minipage}[t]{.44\textwidth}
    \centering
    \includegraphics*[width=\linewidth]{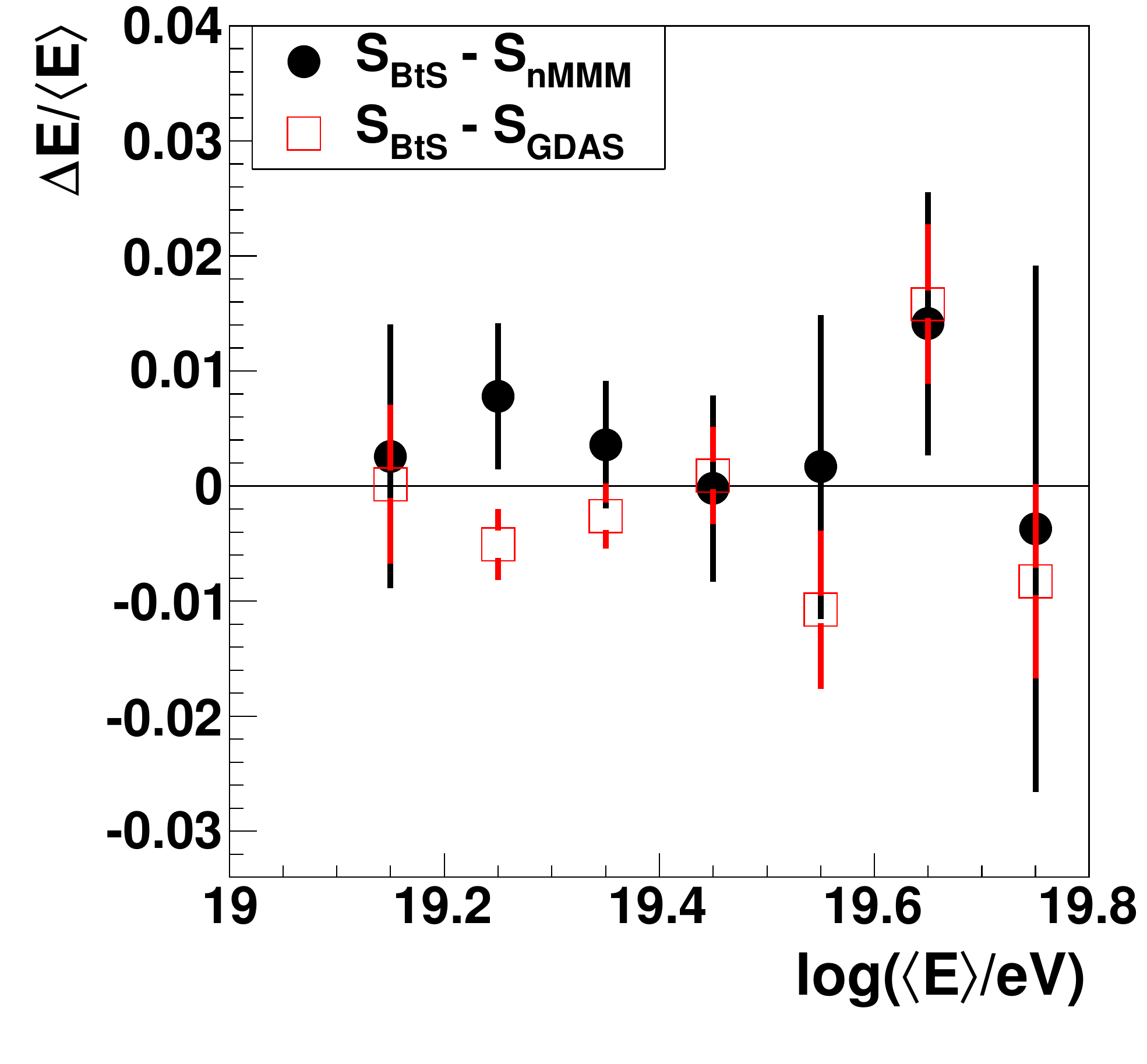}
  \end{minipage}
  \hfill
  \begin{minipage}[t]{.44\textwidth}
    \centering
    \includegraphics*[width=\linewidth]{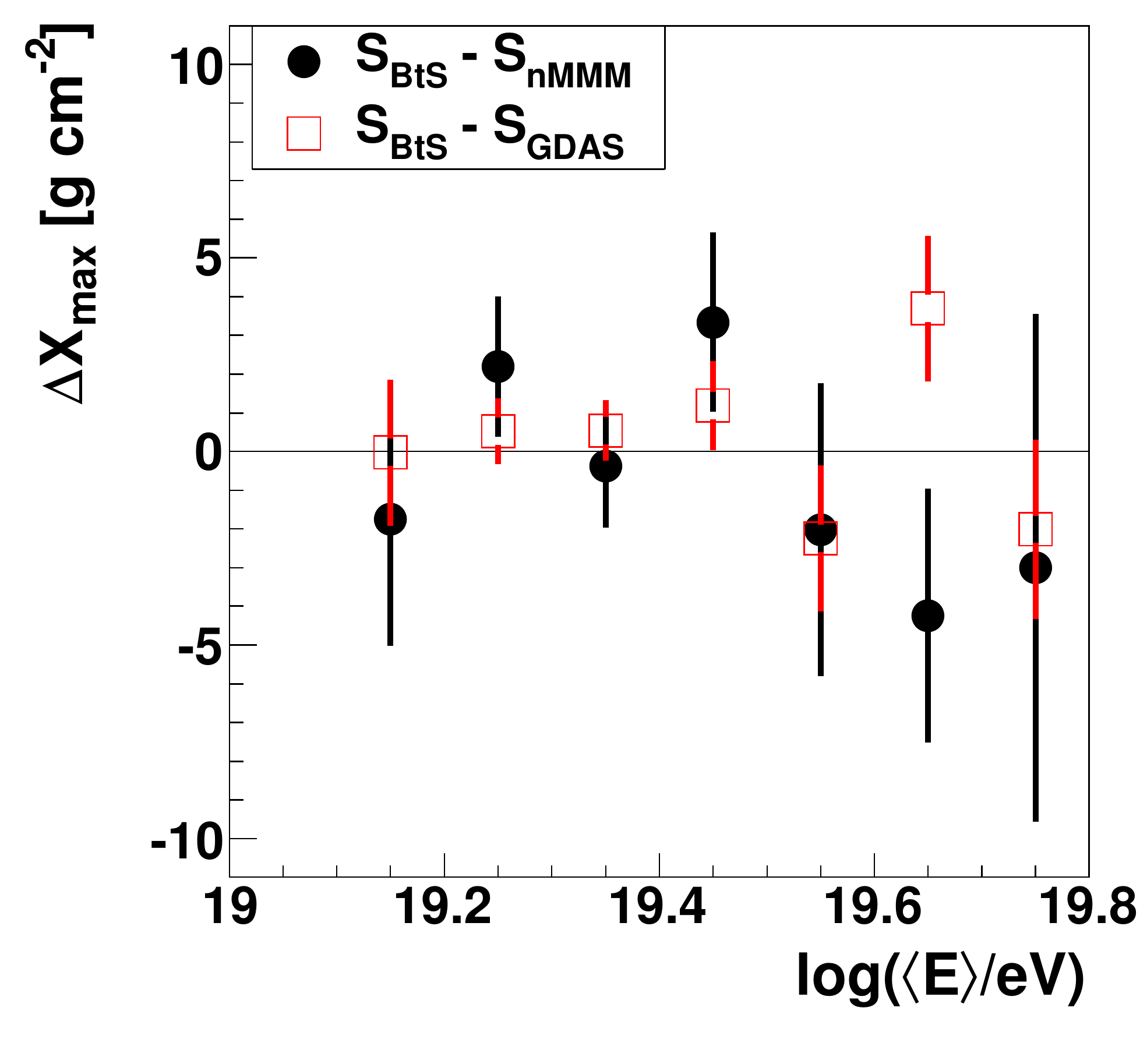}
  \end{minipage}
    \caption{\label{fig:vsE} \slshape
      Differences in reconstructed energy (left) and \xmax (right) vs. energy.
    }
  \vspace{3pt}
  \begin{minipage}[t]{.44\textwidth}
    \centering
    \includegraphics*[width=\linewidth]{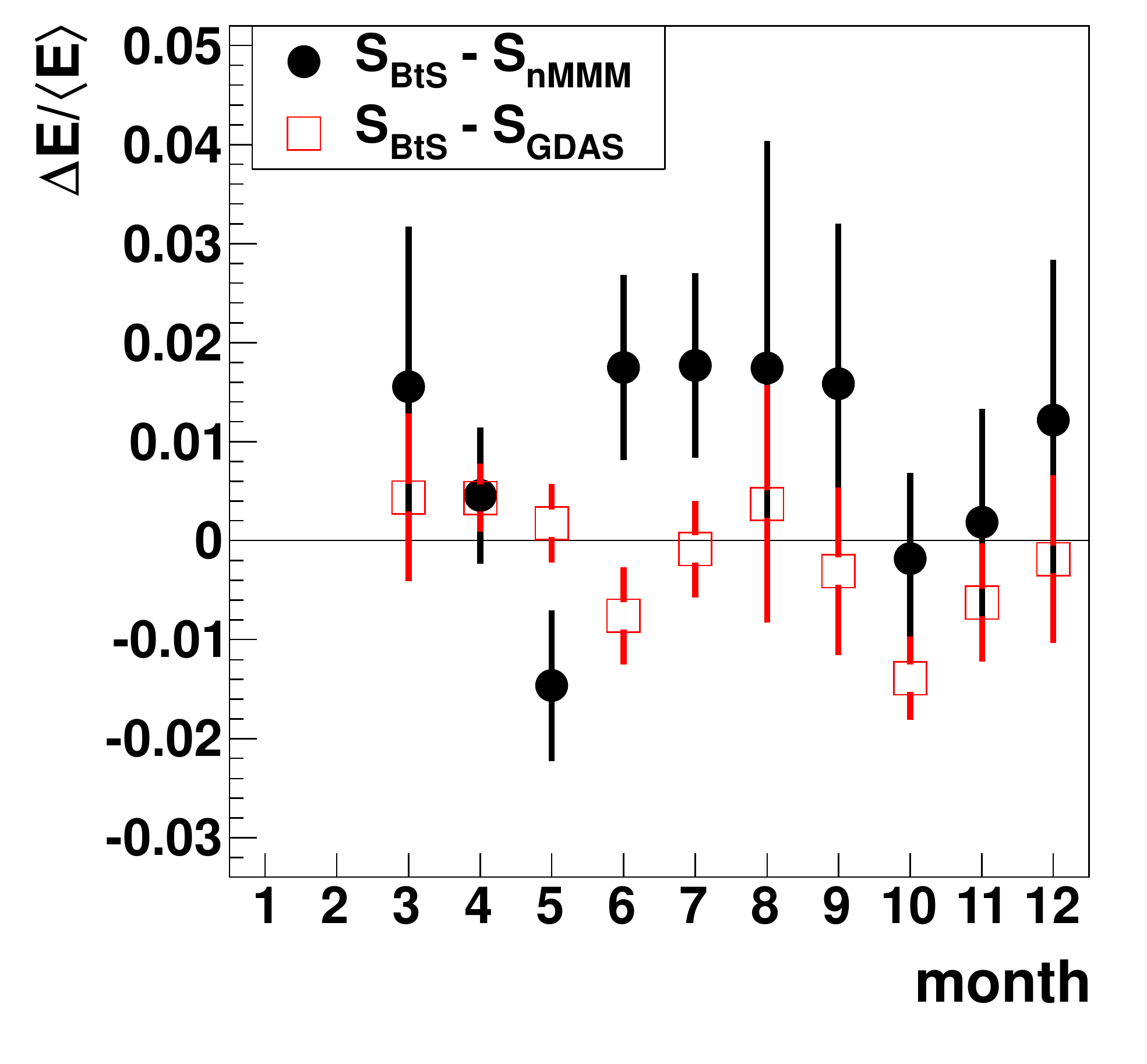}
  \end{minipage}
  \hfill
  \begin{minipage}[t]{.44\textwidth}
    \centering
    \includegraphics*[width=\linewidth]{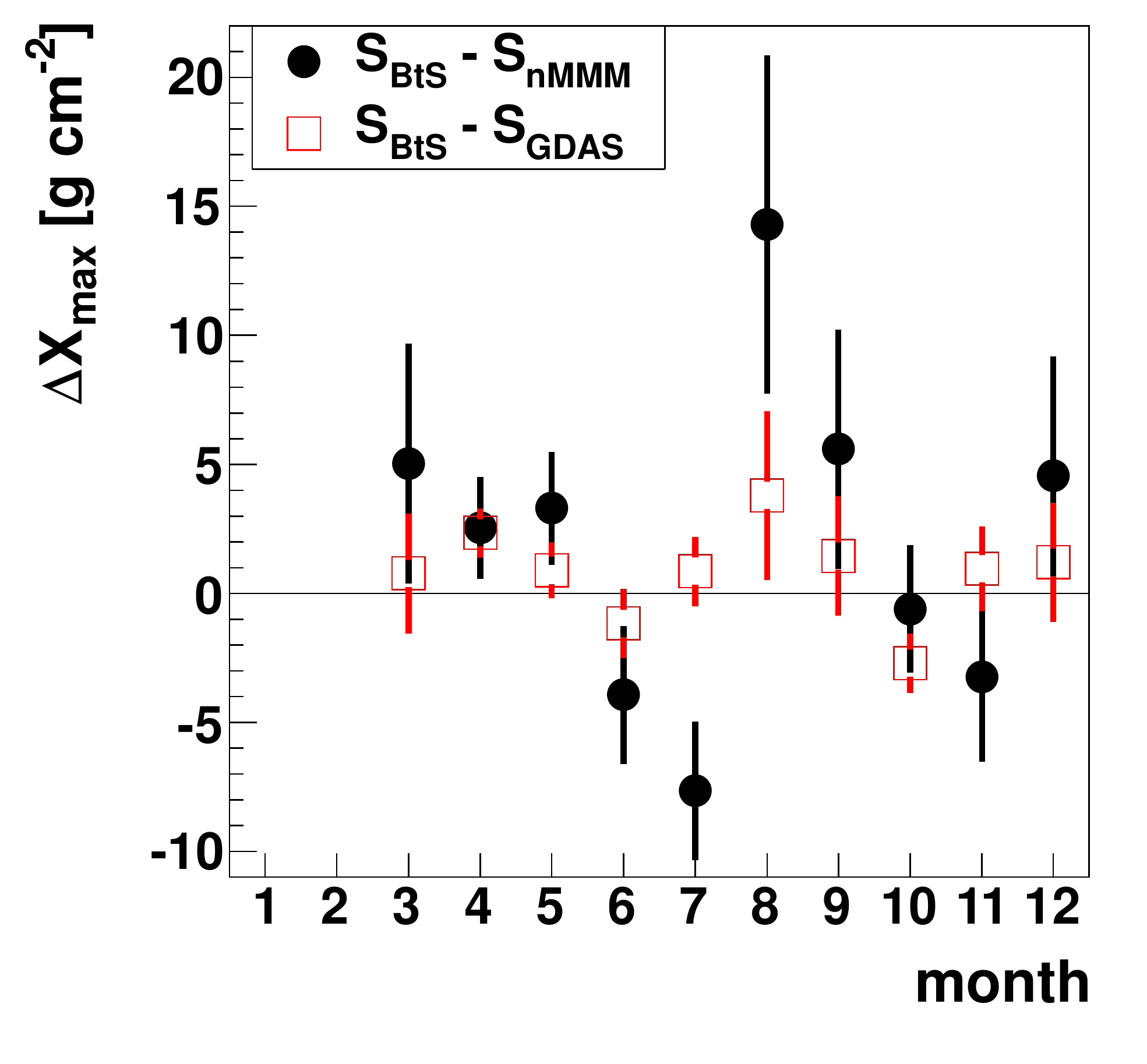}
  \end{minipage}
    \caption{\label{fig:vsMonth} \slshape
      Reconstruction results vs.\ month of year. Left:
      Difference in $E$. Right: Difference in \xmax. There were no balloon
      launches in January or February.
    }
  \vspace{3pt}
  \begin{minipage}[t]{.44\textwidth}
    \centering
    \includegraphics*[width=\linewidth]{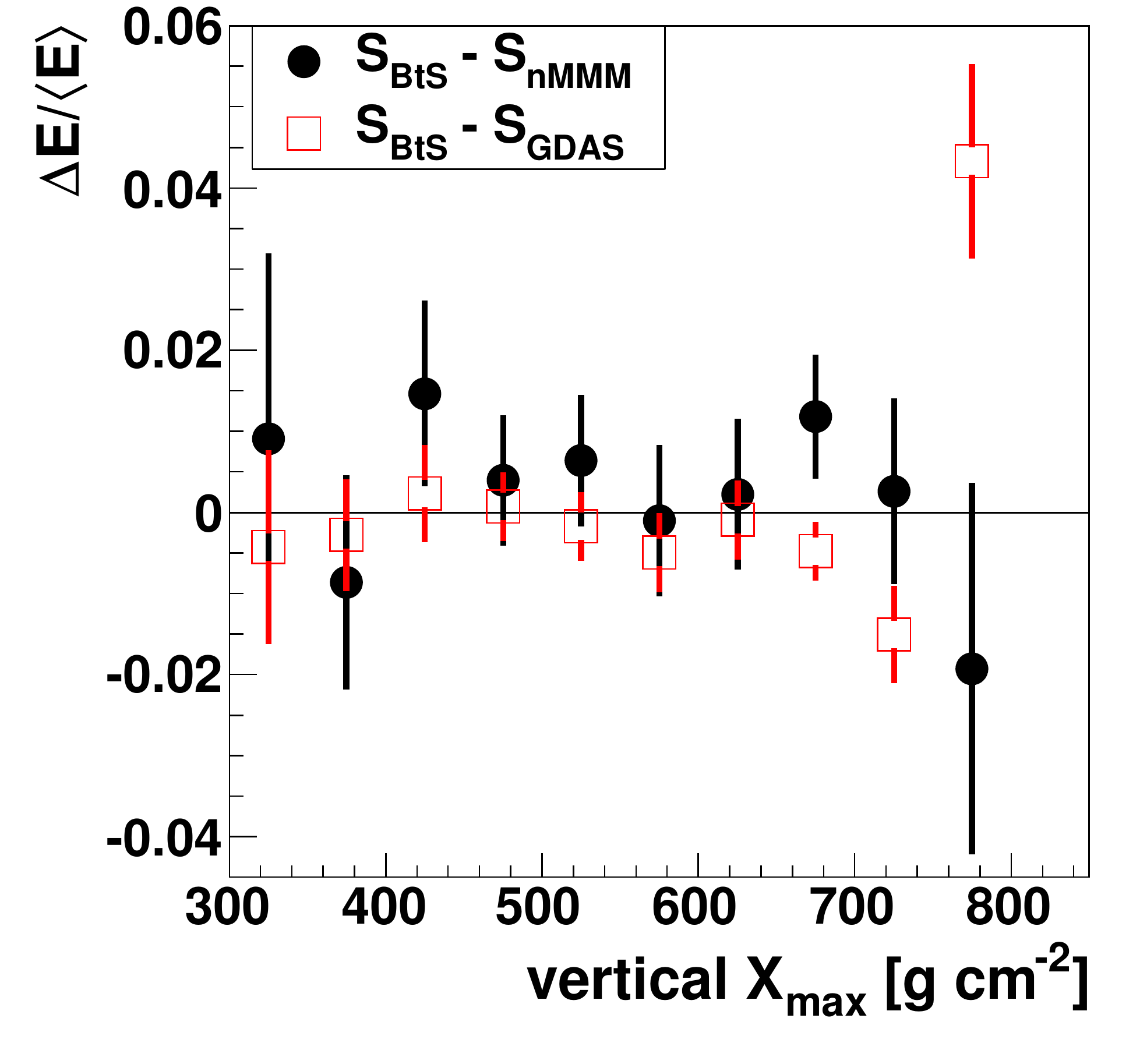}
  \end{minipage}
  \hfill
  \begin{minipage}[t]{.44\textwidth}
    \centering
    \includegraphics*[width=\linewidth]{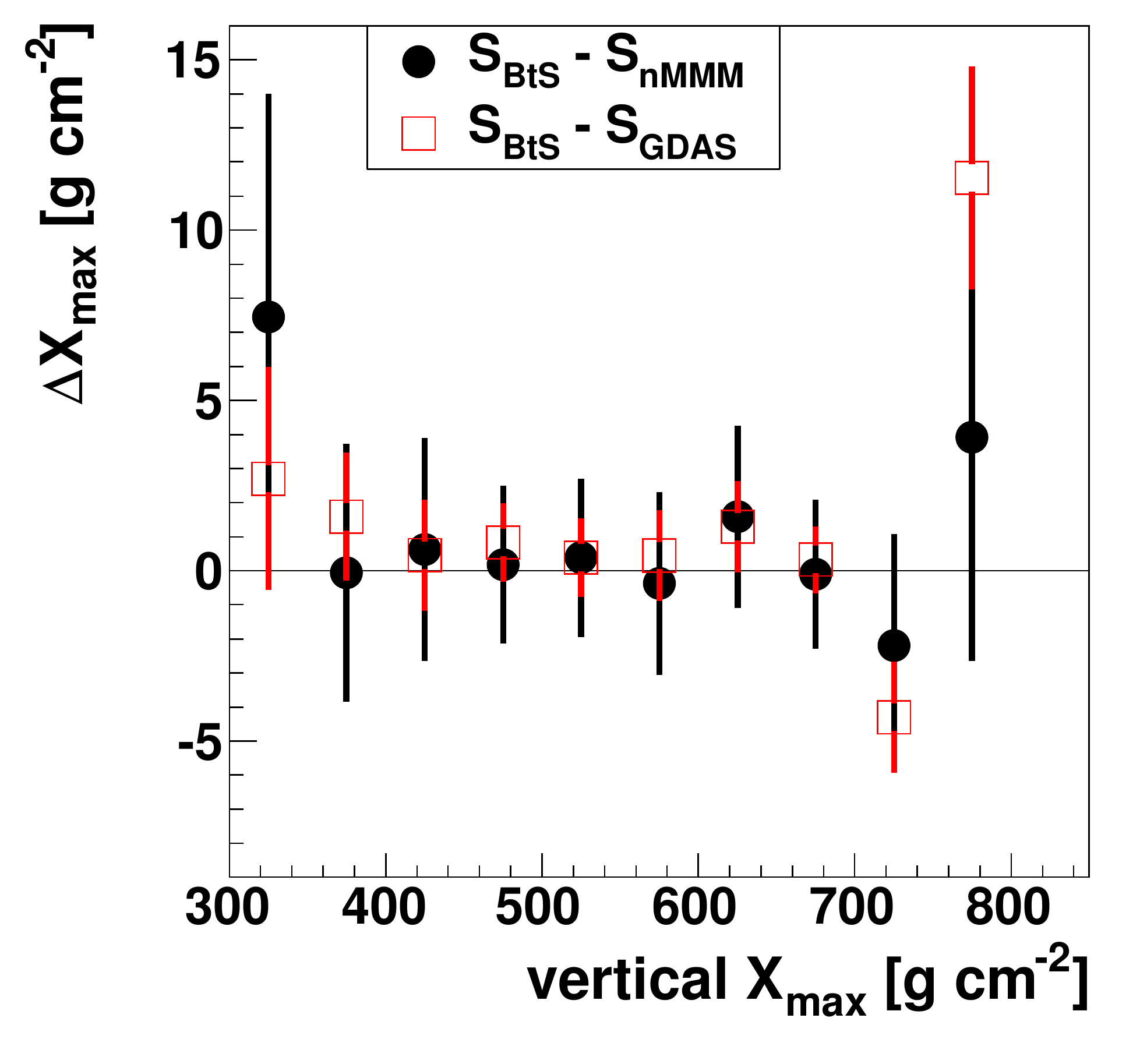}
  \end{minipage}
    \caption{\label{fig:vsXmaxVert} \slshape
      Reconstruction results vs.\ vertical \xmax. Left: Difference
      in $E$. Right: Difference in \xmax.
    }
\end{figure}

In this section, we describe several possible systematic effects in the event
reconstruction:
\begin{enumerate}
  \item Energy dependence of the $\Delta E / \langle E\rangle$ and \relxmax
        distributions;
  \item Seasonal effects;
  \item Dependence on vertical \xmax.
\end{enumerate}
We investigate systematic effects using the differences in the reconstructed
energy and \xmax after using the \bts, \nmmm, and GDAS data in the
reconstruction. The energy dependence of the distributions of \relEavg and
\relxmax is shown in Fig.~\ref{fig:vsE}.  The \sbts-\snmmm comparisons are
plotted with black points, while the \sbts-\sgdas comparisons are plotted with
red squares.  No energy dependence is observed.

Seasonal effects have been investigated by plotting the energy and \xmax
differences according to the calendar month (Fig.~\ref{fig:vsMonth}).  While the
primary energy does not show any signs of seasonal dependence, there are larger
fluctuations in \xmax in particular during the austral winter when using the
\nmmm profiles instead of the \bts profiles in the reconstruction.

The third systematic dependence of interest is vertical \xmax (see
Fig.~\ref{fig:vsXmaxVert}).  Vertical \xmax is the projection of the inclined
shower track to the vertical, which we use to correct for the different
inclination angles of the air showers and establish a clear relationship to the
layering of the atmosphere.  For both $E$ and \xmax, no dependence is obvious.
Note that the entry for vertical \xmax between 750 and 800~\gcmsq corresponds to
only one event, since only one air shower profile has been detected with such a
deep \xmax after applying all quality cuts of the \bts program. This particular
shower entered the Earth's atmosphere quite vertically with a reconstructed
zenith angle of about 12$^\circ$.

Finally, we performed additional searches for any dependence of \relEavg and
\relxmax on fluorescence detector location, on the distance of the core position
from FD telescope, and on some further effects induced by the incoming geometry
of the air shower.  The dependence of \relEavg and \relxmax on these parameters
is negligible in all cases.

\section{{\textit{\textbf Shoot-the-Shower}} Program
\label{sec:sts}}

The purpose of \emph{Shoot-the-Shower} (\sts) is to initiate lidar scans of the
shower-detector plane created by the image of an air shower in an FD telescope.
The motivation is to identify atmospheric non-uniformities -- especially clouds
-- that obscure light from the shower as it propagates to the FD telescopes.
Such non-uniformities may not be present in the hourly atmospheric databases,
and so \sts is intended to supplement the cloud identification performed using
the regular lidar scans.

An hourly cloud coverage below $20\%$ is required for hybrid events to be used
in the analysis of the mass composition and energy spectrum of the cosmic rays
observed at the \pao~\cite{Abraham:2010yv,Abraham:2010mj}.  This cut may still
allow sparse clouds to affect the FD measurements, so one of the main goals of
\sts is to observe showers which pass the cloud coverage cuts but fail the
longitudinal profile cuts.  \sts can be used to verify that the profile quality
cuts are removing showers contaminated by weather effects and not also removing
physically interesting showers from the event sample.  The \sts trigger has
also been adjusted to support the search for anomalous longitudinal profiles
due to hadronic interactions~\cite{Baus:2011icrc}.  These two running modes are
described in the following sections.

\subsection{Performance of \sts}
\label{sec:sts_perform}

Each lidar station contains a steerable telescope and a $150$~\textmu{}J Nd:YLF
laser with a central wavelength of $355$~nm.  The telescope collects
backscattered laser light, and the analysis of the return signal can be
used to infer the presence of aerosols and clouds along the light
path~\cite{BenZvi:2006xb}.  Because the laser wavelength is in the center of
the UV acceptance window of the FD telescopes~\cite{Abraham:2009pm}, the
operation of the lidar must be carefully controlled to avoid triggering the
FD telescopes with scattered laser light.  The implementation of the control
system for \sts is briefly described in Section~\ref{subsec:lidarVeto}.  In
Sections~\ref{subsec:gentrig} and \ref{subsec:dbtrig}, we describe the
general-purpose and anomalous-profile \sts triggers, respectively.

\subsubsection{Full-Site Veto for \sts
\label{subsec:lidarVeto}}

During normal operations, the lidar stations are programmed to observe the
atmosphere above each FD building outside the field of view with an 
automatic scanning mode called
\autoscan.  When an \sts trigger is received, the lidar stations must stop
\autoscan and sweep through the shower-detector plane.  This requires the
lidars to scan inside the field of view of the FD, creating the possibility of
spurious ``self-triggers'', or backscattered laser light triggering a nearby FD
telescope, and ``cross-fires'', or forward-scattered laser light triggering a
telescope on the other side of the SD array.  To avoid spurious triggers, we
have designed the \sts mode to inhibit the FD DAQ for the duration of the \sts
measurement.

\begin{figure}[tb]
  \begin{center}
    \includegraphics*[width=.45\linewidth,clip]{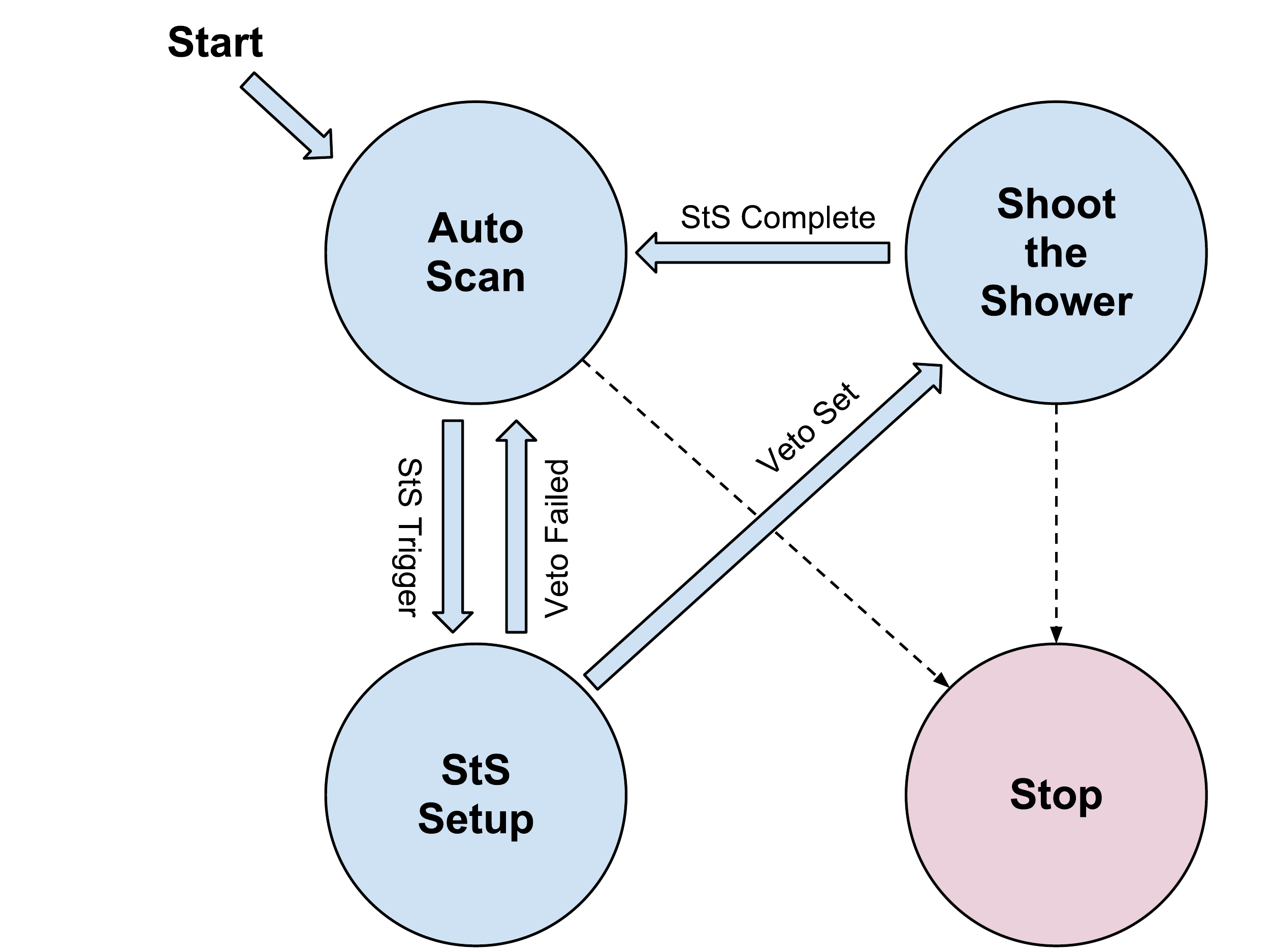}
    \caption{\label{fig:stsStates} \slshape
      Simplified state diagram of the lidar run control.  During normal
      operations, the lidar runs in an automatic scan state (\autoscan).  When
      an \sts trigger is received, the run control software calculates the
      shooting trajectories for all four lidar stations and attempts to veto
      the DAQ of the FD telescopes.  If the veto is set and confirmed, the \sts
      proceeds.  After \sts terminates, the lidar returns to the \autoscan
      state.  The lidar stations will automatically shut down if wind or rain
      exceed safe operating conditions, or if there is a break in network
      communications between the stations and run control in \mal.
    }
  \end{center}
\end{figure}

The implementation of the DAQ veto is shown schematically in the state diagram
in Fig.~\ref{fig:stsStates}.  Online hybrid events are broadcast to the lidar
run control PC running \autoscan in the \mal campus (cf.\
Fig.~\ref{fig:rapidMoniSystem}).  The run control program analyzes the events
for \sts trigger conditions.  When an event passes the triggers, the run
control program calculates a scanning pattern for all four lidar stations and
transmits shooting coordinates to the lidar control PCs at the lidar site.  When the
shooting coordinates are received and automatically confirmed, the run control
program sets a DAQ VETO bit in the GPS servers at each of the FD buildings.
After the veto is set and confirmed by the run control software, the \autoscan
is paused and the lidar stations begin the \sts sweep.

To prevent the lidar run control from deadlocking the FD DAQ, the GPS server
VETO bit is set to revert automatically after four minutes, re-enabling FD data
acquisition.  The lidar \sts coordinates are calculated such that the lidar
telescopes can safely complete the scan during the four-minute shooting window.
If the lidar run control program in \mal loses network communication with the
lidar stations at any time during \autoscan or \sts, the stations will revert
to a partial shutdown mode to prevent uncontrolled laser interference with the
FD telescopes.

\subsubsection{General-Purpose \sts Trigger
\label{subsec:gentrig}}

The first set of quality cuts applied to \sts candidate events, given in
Table~\ref{t:stsCuts}, are designed to include the showers that are likely to
become part of the main high-energy hybrid data set.  However, the cuts are
also loose enough to accept events with unusual ``dips'' and ``spikes'' in the
longitudinal profile.  Such features are typically caused by strong attenuation
and multiple scattering by clouds and aerosol layers.  Including these events
in the \sts sample can allow us to investigate why some longitudinal profiles
do not pass strict profile cuts.  We also considered the possibility that such
events might be recovered for use in the analysis in the future, if the
cloud-affected portions of the longitudinal profiles could be removed.  

\begin{table}[tb]
  \begin{center}
    \begin{tabular}{l l c}
      \toprule
      Quality Cut                 &  & Rejected Events \\
      \midrule
      (1) Field of view           & $X_\text{min}^\text{track}~<~X_\text{max}~<~X_\text{max}^\text{track}$ & 45\% \\
      (2) Gaisser-Hillas (GH) fit & $\chi^2_\text{GH}/n_\text{dof}~<~2.5$, and
                                    $\chi^2_\text{line}~>~\chi^2_\text{GH}$,
                                  & 30\% \\
                                  & $\chi^2_\text{GH}/n_\text{dof}~\geq~2.5$, and
                                    $\chi^2_\text{line}~>~2\chi^2_\text{GH}$
                                  & \\
      (3) Energy uncertainty      & $E~\geq~20~\text{EeV}$, and
                                    $\sigma_E/E~\leq~0.25$,
                                  & 99.7\% \\
                                  & $E~\geq~15~\text{EeV}$, and
                                    $\sigma_E/E~>~0.25$
                                  & \\
      (4) Track length            &  $\Delta X~\geq~300~\text{\gcmsq}$  & 4\% \\
      (5) Local zenith angle      &  $\theta~<~60^\circ$                & 33\% \\
      (6) FD-core distance        & $d_\mr{FD,core}~>~5000~\text{m}$  &  1\% \\
      \bottomrule
    \end{tabular}
    \caption{\label{t:stsCuts}
      A list of quality cuts for the \sts program.  The fraction of rejected events were
      calculated using hybrid data recorded from January 2006 to August 2011.  Note that the
      percentage in each row is given with respect to the previous cut.  The
      event time cut throws out events which are analyzed by the lidar run
      control trigger more than 10~minutes after they were originally detected.
    }
  \end{center}
\end{table}

\begin{figure}[htb]
  \begin{center}
    \includegraphics*[width=.59\linewidth,clip]{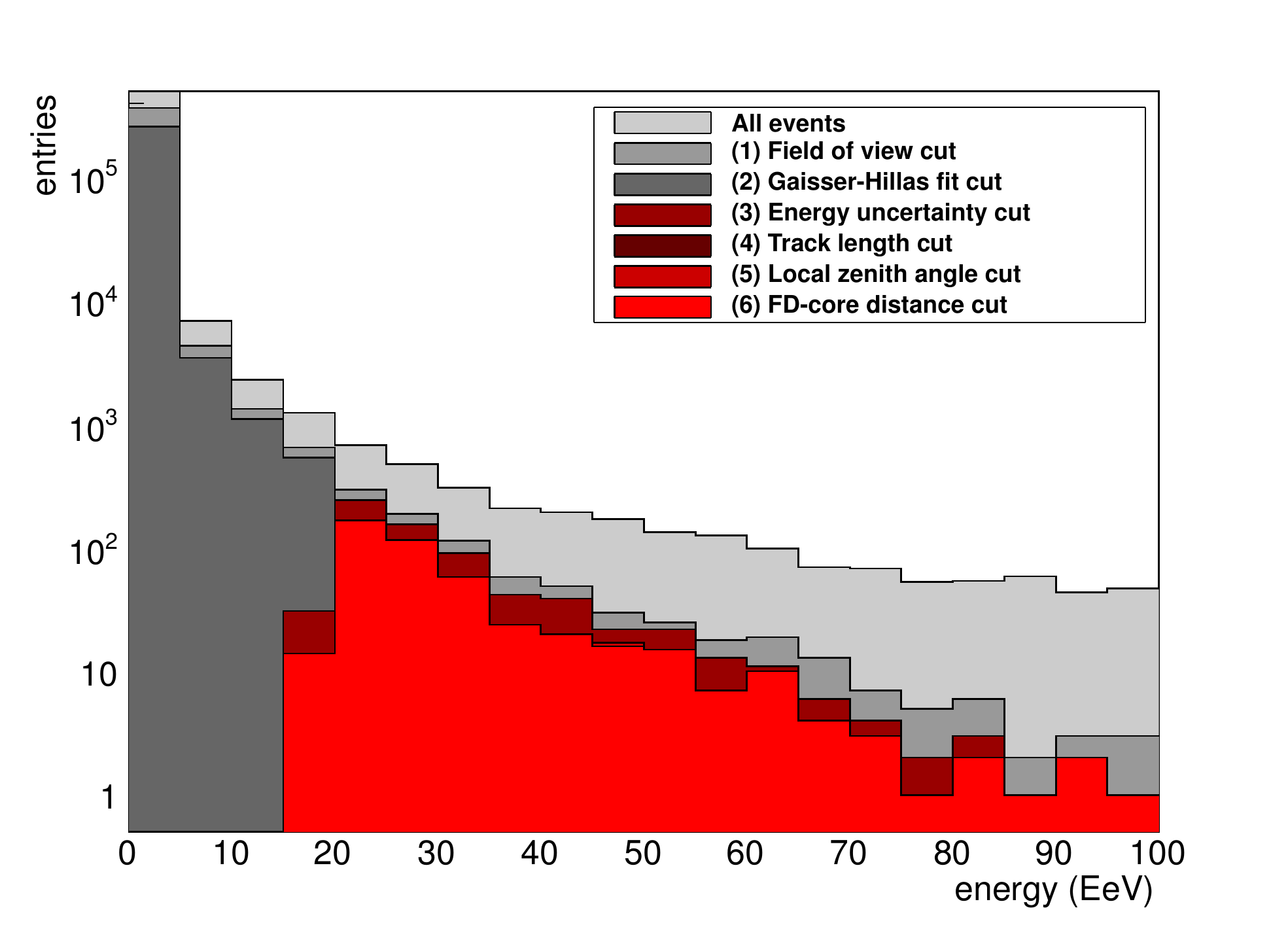}
    \caption{\label{fig:stsCuts} \slshape
      Candidate event distribution for Shoot-the-Shower estimated by using
      hybrid events collected between January 2006 and August 2011. The effect
      of the selection cuts is shown: 99.92\% of the original events are
      discarded.
    }
  \end{center}
\end{figure}

Cuts (1), (2), and (4) in Table~\ref{t:stsCuts} ensure a reliable reconstruction
of the shower energy and the depth of shower maximum.  However, cut (2) on the
shower profile is loose enough to accept events which are not well-described by
a Gaisser-Hillas function.  This includes bumpy events, or profiles with a
strong asymmetry. The purpose of cut (3) on shower energy is to heavily limit
the number of lidar scans: from about two per 17 day FD measurement period during
austral summer up to two per night during winter.  The cut on the zenith angle
(5) rejects overly inclined showers, which are problematic because the scan
path may be so long that a lidar would not finish the \sts within the maximum
veto time window of four minutes.  Finally, cut (6) excludes air showers which
occur at close distances to the FD telescopes, because these do not need
corrections for atmospheric transmission. The effects of the cuts are shown as a
function of energy in Fig.~\ref{fig:stsCuts}.

Events which are analyzed by the lidar run control program are also removed if
the corresponding shower has occurred more than $10$~minutes in the past.  This
is to ensure that the \sts measurements accurately describe the distribution of
clouds and aerosols when the shower was observed.  During times when both the
online reconstruction and the lidar system were operational, no events had to be
rejected because of this criterion.

\subsubsection{Anomalous Profile (``Double-Bump'') \sts Trigger
\label{subsec:dbtrig}}

In March 2011, a second \sts trigger was implemented to aid in the search for
anomalous longitudinal profiles.  This search is motivated by recent
simulations which indicate that in a small fraction of air showers, leading
particles can penetrate deeply into the atmosphere before interacting and
creating another maximum in the shower profile~\cite{Baus:2011icrc}.  The
longitudinal profile of such showers will have two peaks -- hence
``double-bump'' showers -- which can be fit using a sum of two Gaisser-Hillas
functions.  The fit yields two values of \xmax (\xmaxOne and \xmaxTwo) for the
two peaks in the profile. 

\begin{table}[t]
  \begin{center}
    \begin{tabular}{l c c}
      \toprule
      Quality Cut                 &  & Rejected Events \\
      \midrule
      (1) Double GH fit successful  & --- & $1.7\%$ \\
      (2) Field of view cut         & $X_\text{min}^\text{track}~<~X_\text{max,1}~<~X_\text{max}^\text{track}$ & $78.9\%$ \\
                                    & $X_\text{min}^\text{track}~<~X_\text{max,2}~<~X_\text{max}^\text{track}$ & \\
      (3) Double GH $\chi^2$ improvement & $\Delta\chi^2~\leq~-29$ & $99.9\%$ \\
      \bottomrule
    \end{tabular}
    \caption{\label{t:stsCutsDB}
      A list of quality cuts for the \sts ``double-bump'' trigger.  The
      fraction of rejected events were calculated using hybrid events observed
      between 2004 and 2010, after removal of nights contaminated by clouds and
      aerosols.  During online running, the fractions of rejected events are
      substantially larger because clouds and aerosols cannot currently be
      removed in real time.
    }
  \end{center}
\end{table}

The goal of \sts in this analysis is to help discriminate double-peaked
fluorescence profiles caused by hadronic interactions from the much larger
background of double-bump showers caused by scattering due to aerosols and
clouds.  Based on hybrid events observed between January 2004 and December
2010, a set of simple cuts has been developed to select double-bump events for
\sts.  The cuts are listed in Table~\ref{t:stsCutsDB}, and simply require that:
(1) a sum of two Gaisser-Hillas functions can be fit to the longitudinal
profile; (2) both \xmaxOne and \xmaxTwo are within the FD telescope field of
view; and (3) fitting the profile with two Gaisser-Hillas functions results in
a substantial improvement over a single Gaisser-Hillas fit.

We note that the double-bump trigger is not energy-dependent; in principle low
and high-energy showers can give rise to double-peaked profiles.  Therefore, to
prevent the energy cuts described in Section~\ref{subsec:gentrig} from
eliminating double-peaked events, the double-bump trigger has been set to
supersede the standard \sts trigger.  In addition, we note that the cuts in
Table~\ref{t:stsCutsDB} were tuned on a data set in which nights with heavy
cloud and aerosol contamination were removed using the atmospheric databases.
Therefore, the trigger is effective insofar as clouds and aerosols can be
removed in real time.  This issue will be discussed further in
Section~\ref{sec:sts_results}.

\subsection{Results}
\label{sec:sts_results}

\begin{figure}[t]
  \begin{center}
    \includegraphics*[width=.49\linewidth,clip]{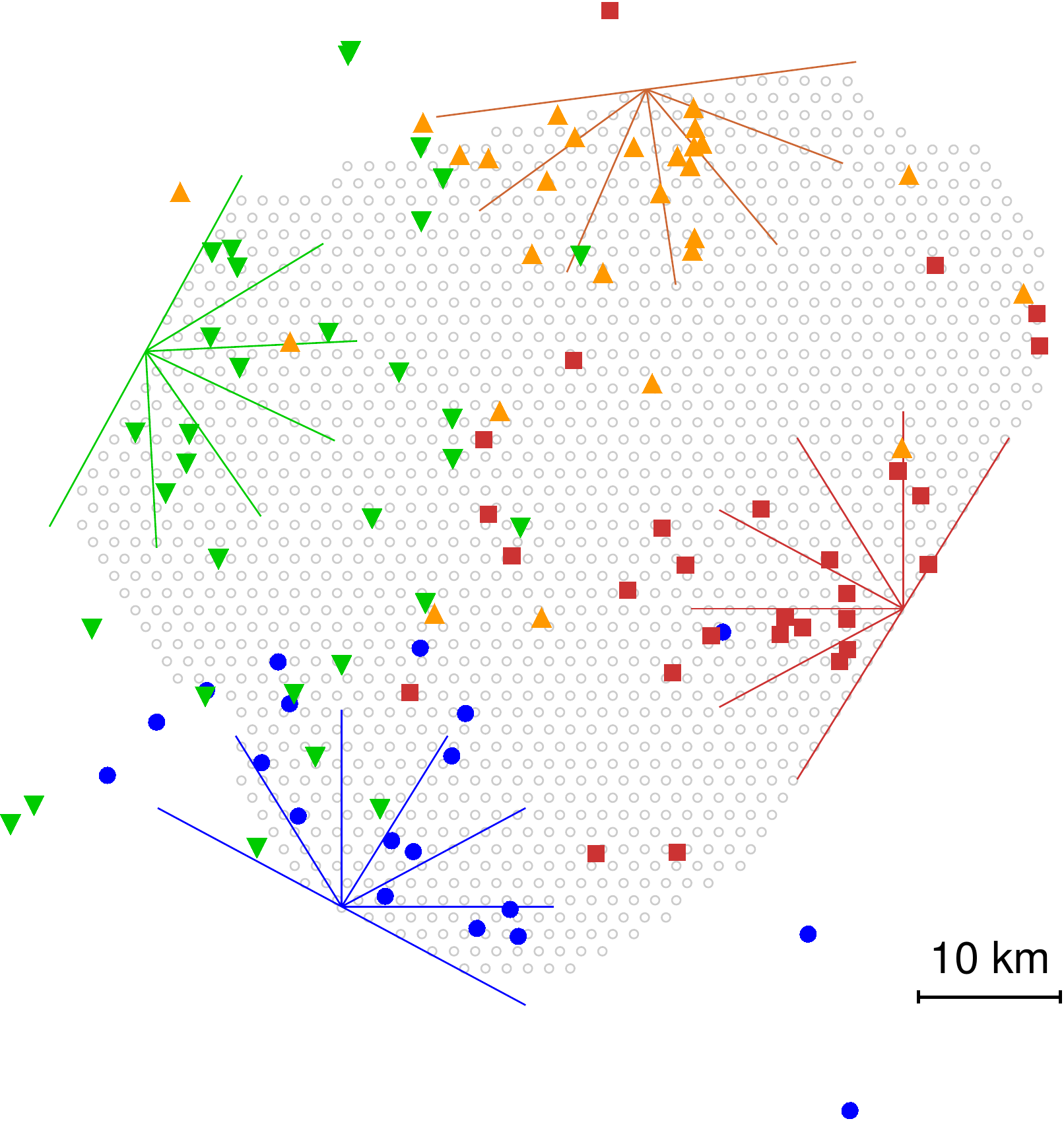}
    \caption[]{\label{fig:stsCores} \slshape
      Core positions of air showers at the array (cf.\
      Fig.~\ref{fig:moniSystems}) which triggered \sts and could be
      reconstructed offline.  The shot statistics are: 20 events at Los Leones
      (circles), 28 at Los Morados (squares), 28 at Loma Amarilla (triangles),
      and 33 at Coihueco (inverted triangles).
    }
  \end{center}
%\end{figure}
%
%\begin{figure}[b]
  \begin{center}
    \includegraphics*[width=0.55\linewidth]{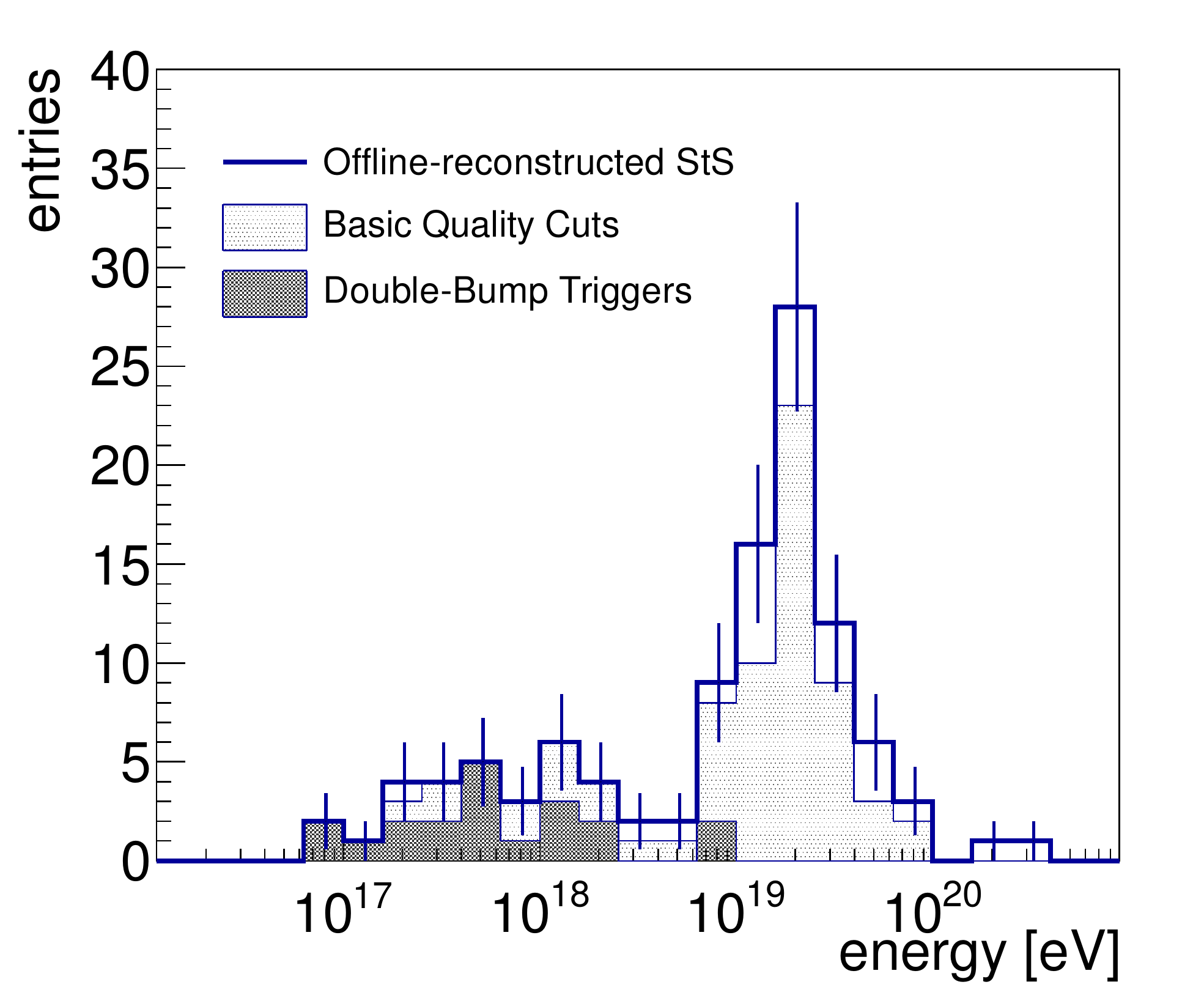}
    \caption{\label{fig:stsEnergy} \slshape
      Distribution of the energies of \sts events, reconstructed offline.  The
      light gray histogram shows events reconstructed as hybrid showers
      (removing ``monocular'' FD events with coincident noise hits in the SD)
      and with \xmax in the field of view of the FD telescopes.  The dark gray
      histogram indicates the events which activated the double-bump trigger.
    }
  \end{center}
\end{figure}

Between January 2009 and October 2011, 112 air showers triggered an \sts scan
and were successfully reconstructed offline.  Of these showers, 58 triggered
telescopes at one FD site; 40 were observed in stereo mode at two FD sites;
eight were observed at three FD sites; and six events were observed at all four
FD sites.  In total this sample comprises 186 individual fluorescence profiles.
The reconstructed ground impact (or core) locations of the 112 air showers are
shown in Fig.~\ref{fig:stsCores}, superimposed on the SD array.  The energies
of the events, reconstructed offline with all available calibration and
atmospheric databases, are shown in Fig.~\ref{fig:stsEnergy}.

Among the notable features in Figs.~\ref{fig:stsCores} and \ref{fig:stsEnergy}
are showers with core locations not contained inside the boundaries of the
array, as well as several very high-energy events ($>10^{20}$~eV).  These
features are due to events in which the SD ``signals'' corresponding to the FD
longitudinal profiles were identified by the \Offline software as
coincident noise.  Therefore, the showers were reconstructed in FD-monocular
mode using no SD data, rather than hybrid mode.  Such events are easily removed from the data by
applying cut~(4) listed in Table~\ref{t:btsCuts}. Applying this cut removes the
showers with cores not contained in the SD array, as well as the high-energy
events in Fig.~\ref{fig:stsEnergy}.  The cut was intentionally omitted from the
\sts trigger because FD-only events are still useful for systematic studies if
they are observed in stereo mode (cf.~\cite{Abbasi:2006dj}).

\subsubsection{Air Shower Analysis using \sts data
  \label{subsec:EASanalys_sts}}

Our main interest in this study is the effect of \sts on the standard hybrid
analyses, and so we cut all events not reconstructed in hybrid mode.  This
reduces the sample to 89 events (146 FD profiles).  We also require that the
remaining FD profiles have a corresponding \sts scan from the lidar station
located at the same FD site.  E.g., an air shower observed at Los Leones must
have an \sts scan from the lidar at Los Leones.  This requirement
reduces the event sample to a final size of 62 events (86 FD profiles).  The
reduced statistics are caused by down time of individual lidar stations during
repair and maintenance periods.

\begin{figure}[htb]
  \begin{center}
    \includegraphics*[width=1.0\linewidth,clip]{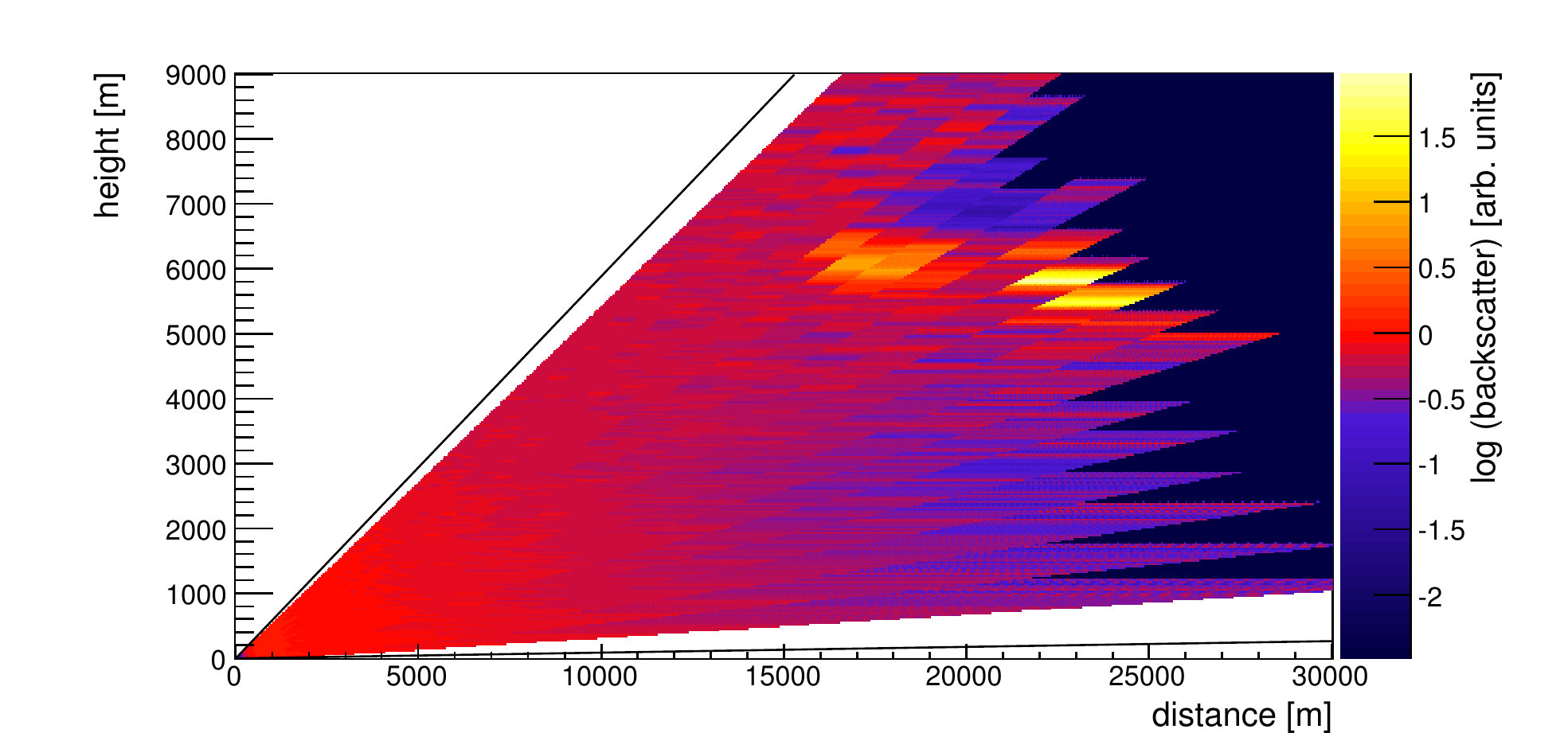}
    \caption{\label{fig:stsscan} \slshape
      Lidar scan from Coihueco after the \sts trigger received at GPS second
      960683560. Altitude is
      measured from the lidar station, which is located about
      1\,700\,m above sea level. The color scale shows the logarithm of the
      backscattered laser light after correction for the $r^{-2}$ decrease in
      the return signal. Bright areas correspond to reflections from clouds.
      For more details about lidar signal processing, see~\cite{BenZvi:2006xb}.
    }
  \end{center}
\end{figure}

The \sts scans have been analyzed in order to find incidental clouds along the
shower path.  An automatic cloud detection algorithm used to estimate cloud
coverage at each FD site~\cite{Tonachini:2010zz} has been adapted for these
scans by implementing a progressive re-binning of the signal trace.  This
allows for an extension of the lidar range up to 30--35\,km, depending on
atmospheric conditions.  An example \sts scan is shown in
Fig.~\ref{fig:stsscan}.

Of the 62 \sts scans, clouds have been detected in 27.  Since the quality cuts
used in most hybrid analyses require periods with less than $20\%$ cloud
coverage, the \sts is particularly important if it identifies clouds during
otherwise clear periods.  It appears that such cases are rather uncommon; of
the 27 \sts scans affected by clouds, only two occurred during periods of low
cloud coverage.  In both cases, the clouds observed were at rather high
altitudes (7 km and 10 km, respectively) and did not appear to affect the
transmission of light between the shower axis and the observing telescopes.

\subsubsection{Analysis of Double-Bump Triggers
  \label{subsec:dbtrigs}}

The \sts double-bump trigger began operating during the FD measurement period of
February-March 2011.  This was a period marked by significant broken cloud
cover nearly every night of the data taking.  As discussed in
Section~\ref{subsec:dbtrig} and Table~\ref{t:stsCutsDB}, the double-bump cuts
were tuned on a cloud-free data set, making the double-bump trigger susceptible
to false positives in the real data where clouds were not removed.  In fact
twenty double-bump triggers were recorded and shot during this measurement period, an
average of 1.3 per night.  Inspection of the data confirmed that all of the
showers were affected by atmospheric scattering.

\begin{figure}[htb]
  \begin{center}
    \includegraphics*[width=0.42\textwidth,clip]{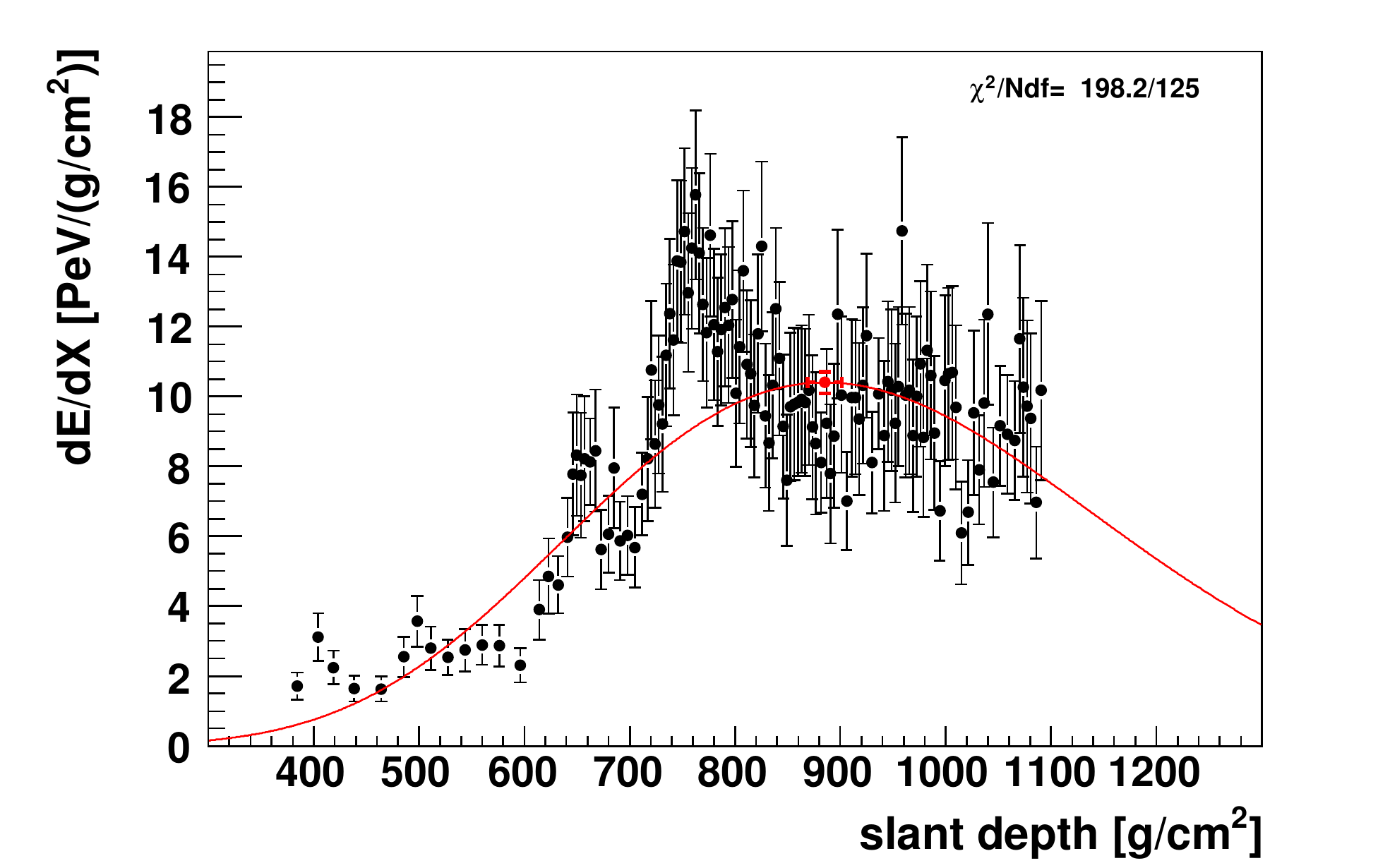}
    \includegraphics*[width=0.57\linewidth,clip]{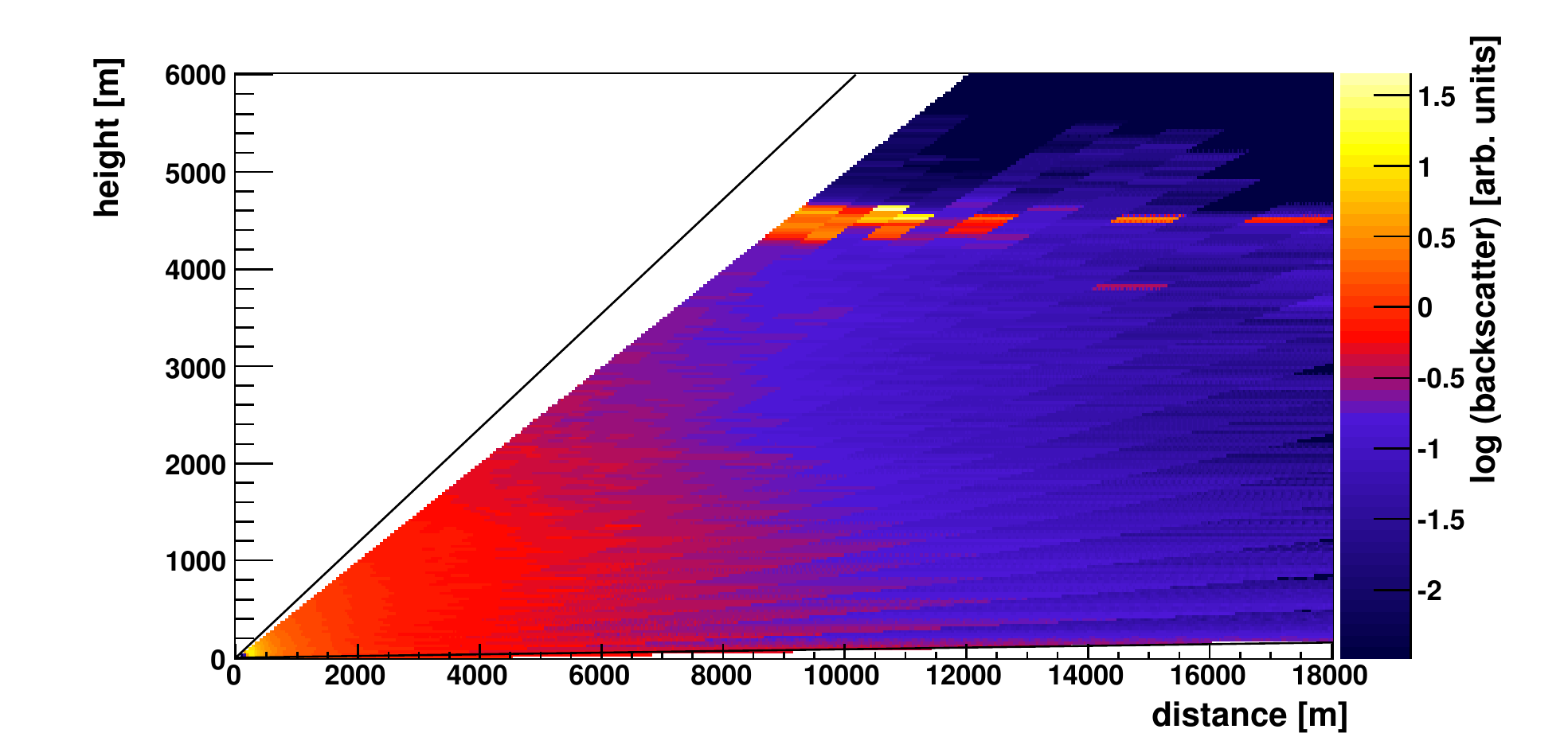}
    \caption{\label{fig:stsEye3} \slshape
      \textsl{Left}: Profile of energy deposit measured at Loma Amarilla,
      5 March 2011,
      showing the shower track and d$E$/d$X$ profile of a double-bump trigger.
      \textsl{Right}: Lidar scan at Loma Amarilla corresponding to this
      ``double-bump'' event.  A layer of broken clouds is clearly visible in
      the light path between the FD and the shower track.}
  \end{center}
\end{figure}

A representative double-bump event, observed at Loma Amarilla, is shown in
Figure~\ref{fig:stsEye3}.  The fluorescence light longitudinal profile
(Fig.~\ref{fig:stsEye3}) exhibits spikes at slant depths of $650$~\gcmsq and
$750$~\gcmsq.  For this shower geometry, the spikes correspond to altitudes of
about 3.6~km and 4.7~km above the FD site Loma Amarilla, which is at an altitude
of about 1.48~km above sea level.  Such spikes are characteristic of
multiple-scattering of light as the air shower enters a cloud layer.  Inspection
of the \sts profile recorded by the Loma Amarilla lidar station confirms the
presence of a reflective cloud layer in the shower-detector plane at the
expected altitude that is responsible for the spikes.

Unfortunately, an analysis of the first double-bump profiles reveals that all
of the triggers were clearly spurious events created by obscuring clouds. In
order to avoid wasting valuable time shooting low-energy showers affected by
clouds, it is clear that some kind of real-time identification of coverage will
be necessary.  This may already be possible with existing lidar data.  The
cloud-detection algorithm used on the \autoscan data and described
in~\cite{Tonachini:2010zz} is both fast and robust, and could be applied
online.  An online cloud coverage estimate could then be used to suppress \sts
after double-bump triggers whenever the coverage was over some minimum
threshold, such as $20\%$.  The real-time identification of cloud coverage and
its application to \sts is currently under study.

\section{Rapid Monitoring with FRAM
\label{sec:fram}}

The (F/Ph)otometric Robotic Atmospheric Monitor, or FRAM telescope, is capable
of carrying out rapid monitoring observations.  FRAM is operated as a passive
scanner -- i.\,e., observations do not require the use of lasers or light
flashers -- and so it does not introduce any dead time into the FD data
acquisition. Therefore, unlike the lidars, there are no limitations on the use
of FRAM in the rapid monitoring program.  FRAM can be programmed to scan the
shower-detector plane of interesting showers and record CCD images along the
plane.  Since the FRAM telescope is located close to the FD building at Los
Leones, its use in the rapid monitoring program is limited only to hybrid air
showers observed from Los Leones.  From this location, sequences of CCD images
are produced along the shower-detector planes of observed cosmic ray events.

\begin{figure}[tb]
  \begin{center}
    \includegraphics*[width=.59\linewidth,clip]{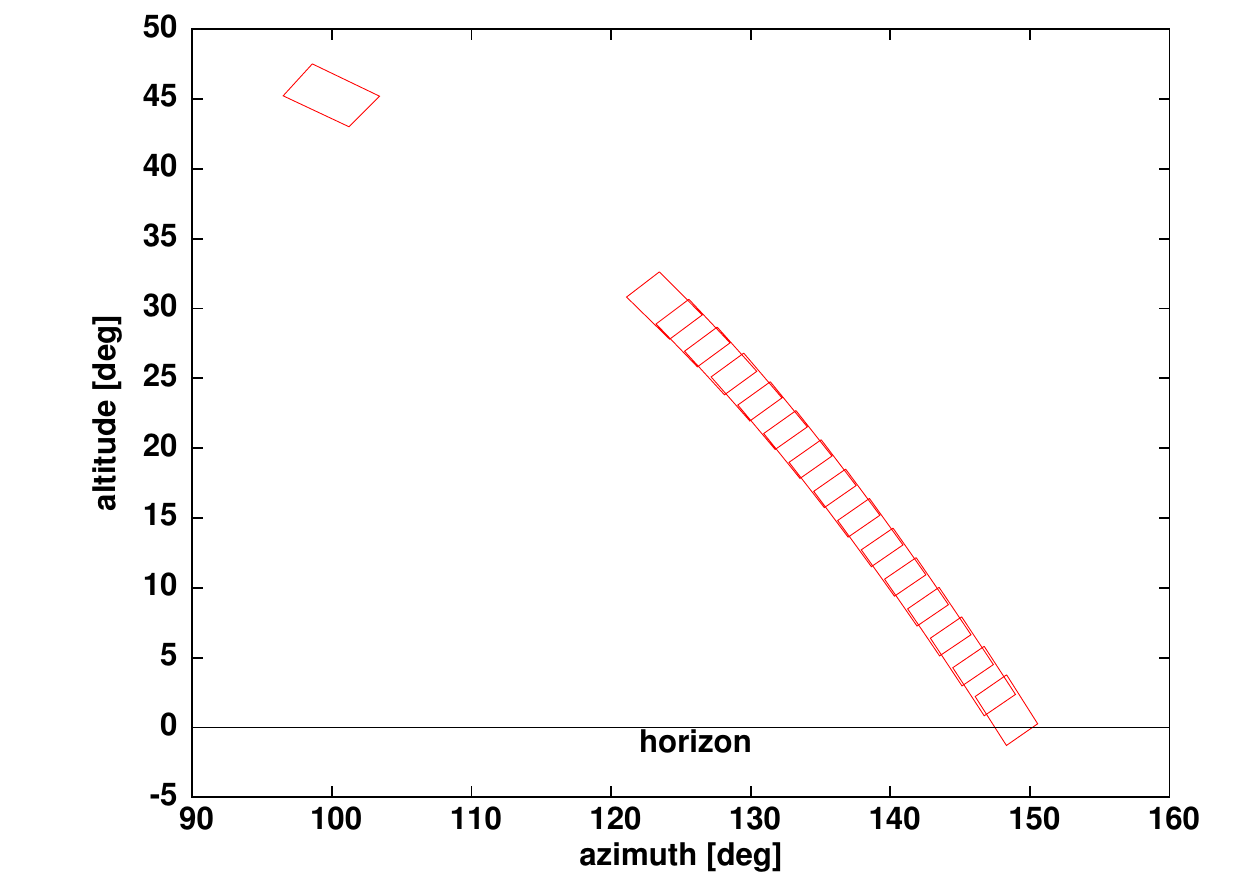}
    \caption{\label{fig:fram_trajectory} \slshape
      Example of a shower track sampled by individual exposures of the CCD
      camera. The red rectangles show the approximate fields of view of the wide
      field camera images.  The distortions of the shapes of the fields of
      view are due to the projection onto rectangular coordinates.  The
      isolated rectangle in the upper left corner is the first image in the
      sequence, taken along the arrival direction of the cosmic ray shower.
    }
  \end{center}
\end{figure}

The FRAM telescope is equipped with two CCD cameras.  A
wide-field\footnote{Finger Lake Instrumentation (FLI) MaxCam CM8 with Carl Zeiss
Sonnar 200~mm $f/2.8$ telephoto lens.} (WF) camera is used to measure the
atmospheric extinction along the shower-detector plane, and a
narrow-field\footnote{Since June 2010, we have used the Moravian Instruments CCD
camera G2.  Before June 2010, we used another FLI MaxCam CM8.  Both cameras have
been situated at the focus of the main 30 cm-diameter telescope.} (NF) camera is
used to calibrate the WF images.  The field of view of the WF camera is
$240^{\prime}$ ($4^\circ$) in azimuth (aligned with right ascension) and
$160^{\prime}$ ($2.67^\circ$) in elevation (aligned with declination).  Hence, a
shower traversing the whole field of view of an FD telescope is typically
covered by 10 to 20~CCD images (Fig.~\ref{fig:fram_trajectory}).  The first
image of the sequence is oriented along the axis of the cosmic ray air shower to
search for optical transients that could be associated with event, assuming the
primary cosmic ray was a neutral particle.  The field of view of the NF camera
is about $25^{\prime}$ ($0.4^\circ$) in right ascension and $17^{\prime}$
($0.3^\circ$) in declination, and is centered within the WF camera image.

The images recorded with FRAM are analyzed automatically. First, stars in the
image are identified and their observed magnitudes are compared with values from
a catalog, allowing computation of the measured atmospheric extinction along the
line of sight to each star.  As there are typically several hundred identifiable
stars in each image, it is possible to monitor the extinction along the shower
track with high angular resolution.  Variations of the resulting extinction
coefficient reveal the presence of clouds or prominent aerosol layers.

Currently, the rapid monitoring observations of FRAM are carried out using the
Johnson-Cousins B filter (central wavelength 435~nm; FWHM 100~nm) of the UBVRI
set of astronomical filters~\cite{Bessel:1990}.  Both cameras are equipped with
the same set of filters. For the comparison with FD observations, the
Johnson-Cousins U filter (central wavelength 360~nm; FWHM 64~nm) would be more
suitable, but the transmission of the WF camera of the FRAM telescope decreases
very quickly below 400~nm, and in the U filter we typically detect only several
stars, too few for a successful computation of the extinction along the
shower-detector plane.

Following astronomical convention, the light extinction is obtained by
calculating the difference between the measured magnitude $m_\text{obs}$ of
observed stars and the tabulated magnitude $m_\text{tab}$ available in star
catalogs~\cite{Hog:2000}.  This difference depends on the optical path length
through the atmosphere, known in astronomy as the airmass $AM$.  The airmass is
a zenith-dependent quantity similar to slant depth that is expressed with
respect to the total atmospheric overburden observed along a vertical path
between sea level and the top of the atmosphere.  From these quantities, the
extinction can be expressed as an extinction coefficient
$k=(m_\text{obs}-m_\text{tab})/AM$.  The extinction coefficient can then be
transformed to optical depth using the expression\footnote{Since the magnitude
system is one of relative brightness, a star that has a magnitude five times
greater than that of another star will have a light intensity 100 times
greater.}
\begin{equation}
  \label{eq:Tau}
  \tau = \frac{\sqrt[\leftroot{4}\uproot{2}5]{100}}{e}\,k = 0.924\,k.
\end{equation}

The relatively straightforward analysis of the CCD images is an advantage of the
FRAM rapid monitoring program, because the analysis can be almost completely
automated.  Currently, the analysis is carried out offline by the telescope
operator, but a fully automated analysis of the FRAM rapid monitoring data is
foreseen in the near future.

\subsection{FRAM cuts for rapid monitoring}
\label{sec:fram_cuts}

The passive nature of the FRAM observations allows for few constraints in the
selection of suitable events relative to the cuts applied for \sts (see
Sec.~\ref{sec:sts_perform}). Currently, three sets of cuts are applied to focus
on different types of analyses.  The cut parameters can be modified
interactively using the RTS2 control software of the
telescope~\cite{Kubanek:2006}, even on very short time scales such as one night.

The three sets of cuts are shown in Table~\ref{tab:FRAMcuts}.  Set 1 has been
chosen to match the standard selection of high-quality hybrid events used in the
analysis of the energy spectrum~\cite{Abraham:2010mj}.  Set 2 applies the
selection criteria used to identify very deeply-penetrating photon-induced
showers, closely matching the cuts used to estimate the photon upper limit with
hybrid events~\cite{Scherini:2009}.  Finally, the third set of cuts is a relaxed
version of the standard quality selection.  It is designed to exploit the full
capability of the FRAM passive monitor, i.\,e., to perform rapid follow-up
monitoring of a very large set of events without interrupting the FD DAQ.  The
only criterion that is more restrictive in set 3 than in set 1 is the number of
triggered PMTs of the FD camera.  Without this, set 3 would generate too many
triggers.  The higher number of PMTs ensures that showers are selected with long
tracks that might be better covered by the series of CCD images. The three types
of triggers are combined in a logical \textsf{OR} so that a shower fulfilling
any one set will trigger rapid monitoring with FRAM.
\begin{table}[tbp]
  \begin{center}
    \begin{tabular}{rlccc}
      \toprule
      \multicolumn{2}{c}{Criteria} & \multicolumn{3}{c}{Set of Cuts} \\
      & & 1 & 2 & 3 \\
      & & (standard) & (photon showers) & (relaxed standard) \\
      \midrule
      1. & Reconstructed $E$ & $\ge 10^{18}$~eV &  $\ge 10^{18}$~eV &  $\ge 10^{18}$~eV \\
      2. & Relative uncertainty $\sigma_E/E$ & $\le 0.2$ & --- & $\le 0.4$ \\
      3. & Uncertainty of \xmax & $\le 40$~\gcmsq & --- & $\le 80$~\gcmsq \\
      4. & Triggered PMTs & --- & $\ge 6$ & $\ge 10$ \\
      5. & Profile quality: $\chi^2_\text{GH}/\text{n}_\text{dof}$ & $<$~2.5 & $<$~6 & $<$~10 \\
      6. & Comparison of GH and linear fit: & & & \\
      & ~~~~$\chi^2_\text{lin} - \chi^2_\text{GH}$ & $> 4$ & --- & $> 1$ \\
      & ~~~~$\chi^2_\text{GH}/\chi^2_\text{lin}$ & --- & $< 0.9$ & --- \\
      7. & Zenith angle $\theta$ of air shower & $< 70^{\circ}$ & --- & --- \\
      8. & FD telescope-shower viewing angle & --- & $>15^\circ$ & --- \\
      9. & Hottest station-core distance & $< 2000$~m & $< 1500$~m & $< 2000$~m \\
      10. & Time elapsed since shower arrival & $\le$~600~s & --- & --- \\
      \bottomrule
    \end{tabular}
    \caption{\label{tab:FRAMcuts}
      Three sets of cuts for the FRAM rapid monitoring program.
      Since March 2011, the energy threshold for all three sets has been lowered to
      $E \ge 3 \times 10^{17}$ eV.
    }
  \end{center}
\end{table}

In March 2011, the FRAM trigger was updated to include a search for anomalous
``double-bump'' showers.  As with lidar \sts, the double-bump trigger includes
an additional cut on the improvement in the fit of the shower profile when using
two Gaisser-Hillas functions.  To accommodate such observations, we lowered the
energy threshold for the FRAM rapid monitoring triggers to $3\times10^{17}$~eV
in March 2011.  However, a dedicated trigger filter for anomalous events is
currently being implemented.  We describe the search for anomalous shower
profiles in Section~\ref{sec:fram_results}.

\subsection{Performance of the FRAM telescope}
\label{sec:fram_performance}

The FRAM rapid monitoring program has operated since November 2009. For the
first two months, it was running in a test mode. The cuts given in
Table~\ref{tab:FRAMcuts} have been applied since January 2010.

To successfully analyze a CCD image, it is necessary to obtain its astrometry.
This means identifying the stars recorded in the image using a catalog in order
to obtain their exact positions and catalog magnitudes. The shower track is
scanned by both CCD cameras at the telescope, but only the wide-field WF camera
fully covers the shower track without gaps. Therefore, the data from the
narrow-field NF camera are only used to cross-check values from the WF camera.
For analysis of the shower-detector plane, we require at least 5 images from the
WF camera with astrometry.

Between January 2010 and the end of July 2011, the FRAM telescope received
1\,038 triggers for rapid monitoring fulfilling at least one of the three set of
cuts described in Table~\ref{tab:FRAMcuts}.  Of those, 586 were successfully
observed --~i.\,e., at least 5~WF camera images with astrometry were obtained~--
with 242 of showers observed in 2010 and 344 in 2011\footnote{The observations
of showers in periods of bad weather with many clouds do not have a sufficient
number of images with astrometry. The rate of successful observations was lower
in 2010 due to telescope software and hardware issues, and it has increased in
2011.}. The much higher event rate in 2011 was achieved after an upgrade of the
telescope control software in March 2011.  To allow for more frequent
observations, the energy threshold for the triggers in Table~\ref{tab:FRAMcuts}
was lowered to $3 \times 10^{17}$ eV.

\begin{figure}[tb]
  \begin{center}
   \includegraphics*[width=.65\linewidth,clip]{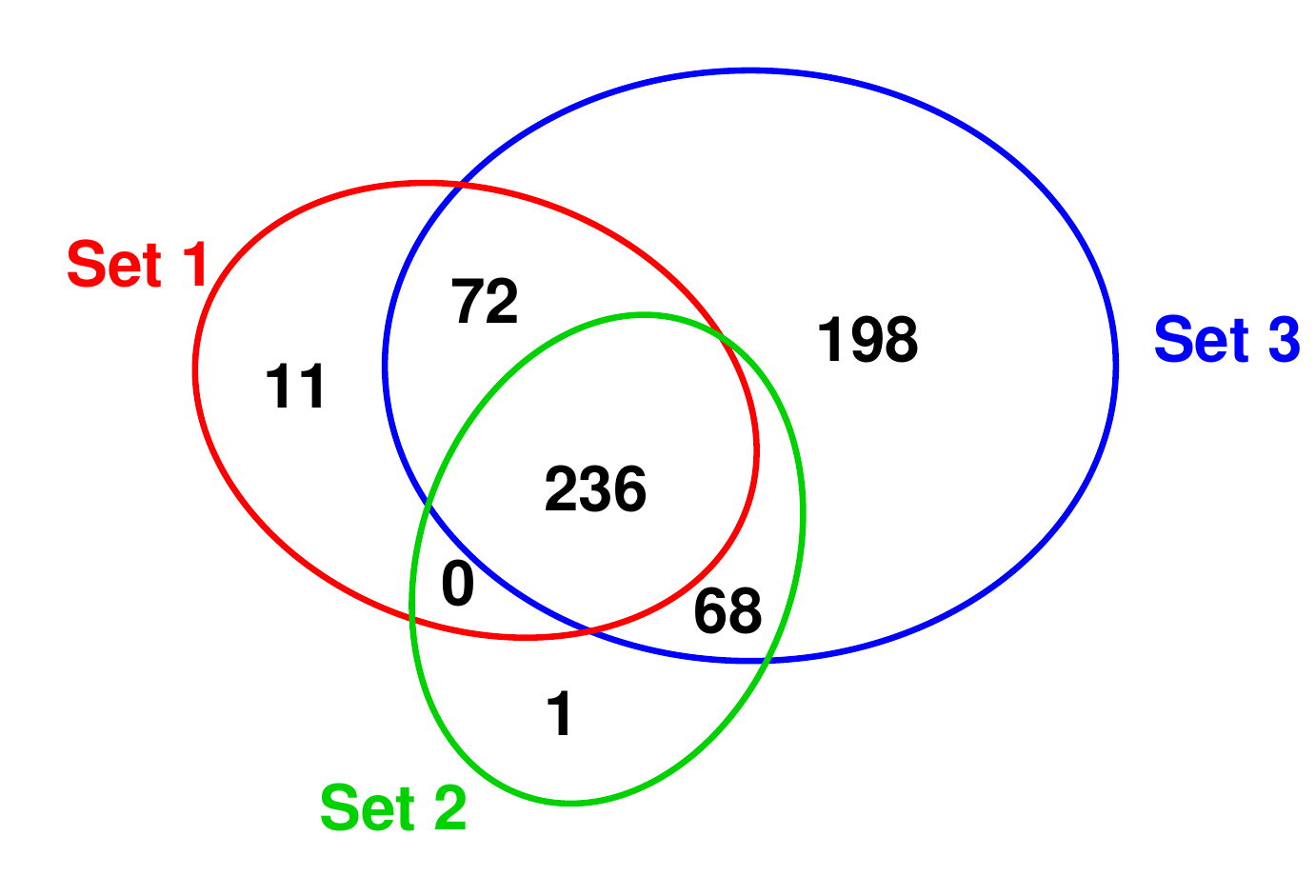}
    \caption[]{\label{fig:fram_cut_distrib} \slshape
    Distribution of successful rapid monitoring observations by the FRAM
    telescope according to different sets of cuts as given in
    Table~\ref{tab:FRAMcuts}.
    }
  \end{center}
\end{figure}

The overwhelming majority of the showers, 574 out of 586, passed the relaxed
trigger condition (set 3).  The standard trigger (set 1) was more restrictive,
passing 319 events, as was the photon trigger (set 2), which passed 305 events.
The statistics of each trigger filter and the overlap between them are shown in
Fig.~\ref{fig:fram_cut_distrib}.  The overlap between the filters is relatively
large; 236 showers passed all three trigger conditions.  Only 11 events were
exclusively passed by the standard cuts, and just one event was passed by the
photon cut alone.  However, the photon trigger was affected by a bug before July
2011 that caused events to be lost, so we note that the passing statistics from
set 2 are artificially low with respect to the other two filters.

\subsection{Results
\label{sec:fram_results}}

\begin{figure}[htb]
  \begin{center}
    \includegraphics*[width=.49\linewidth,clip]{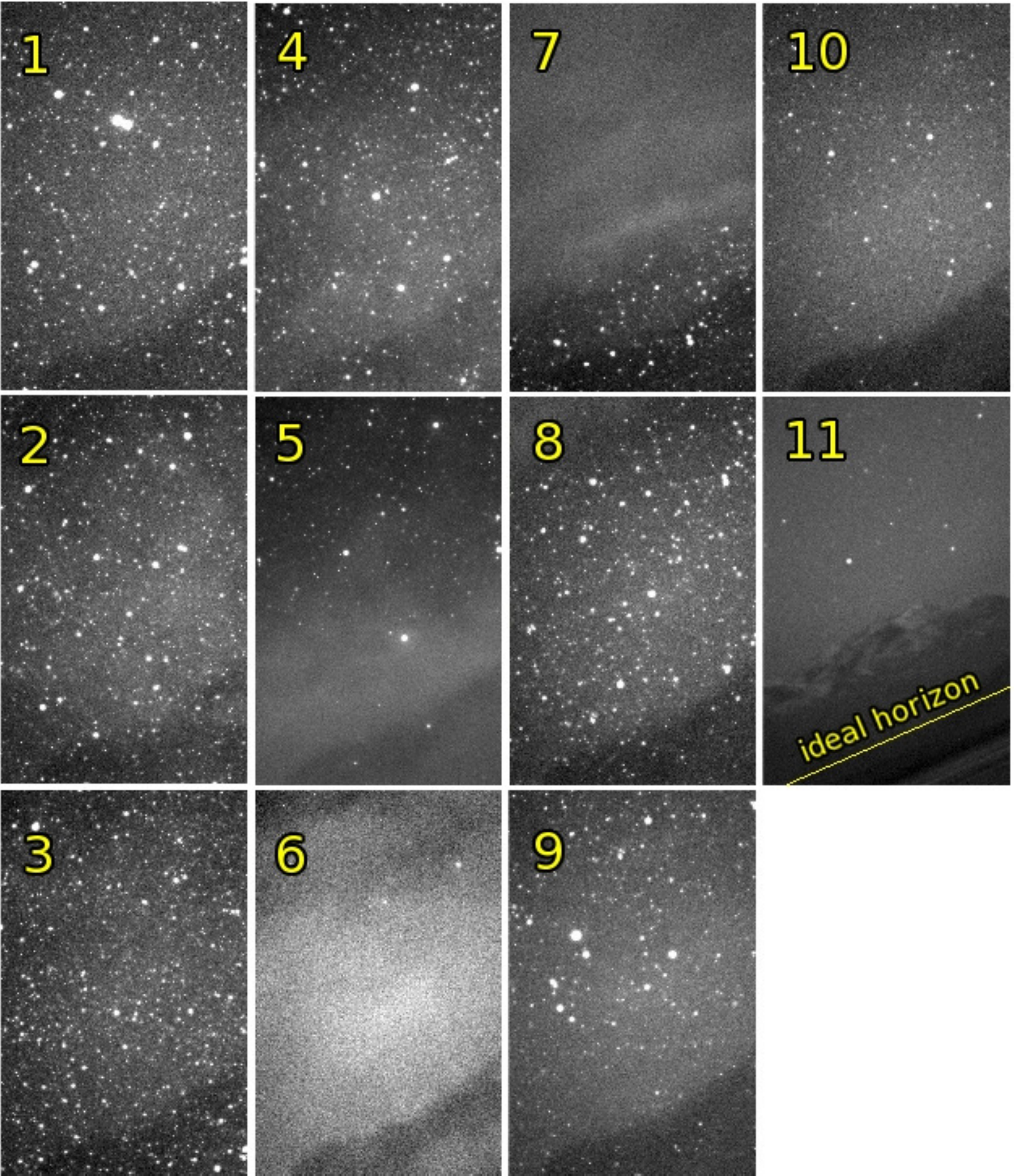}
    \hfill
    \includegraphics*[width=.47\linewidth,clip]{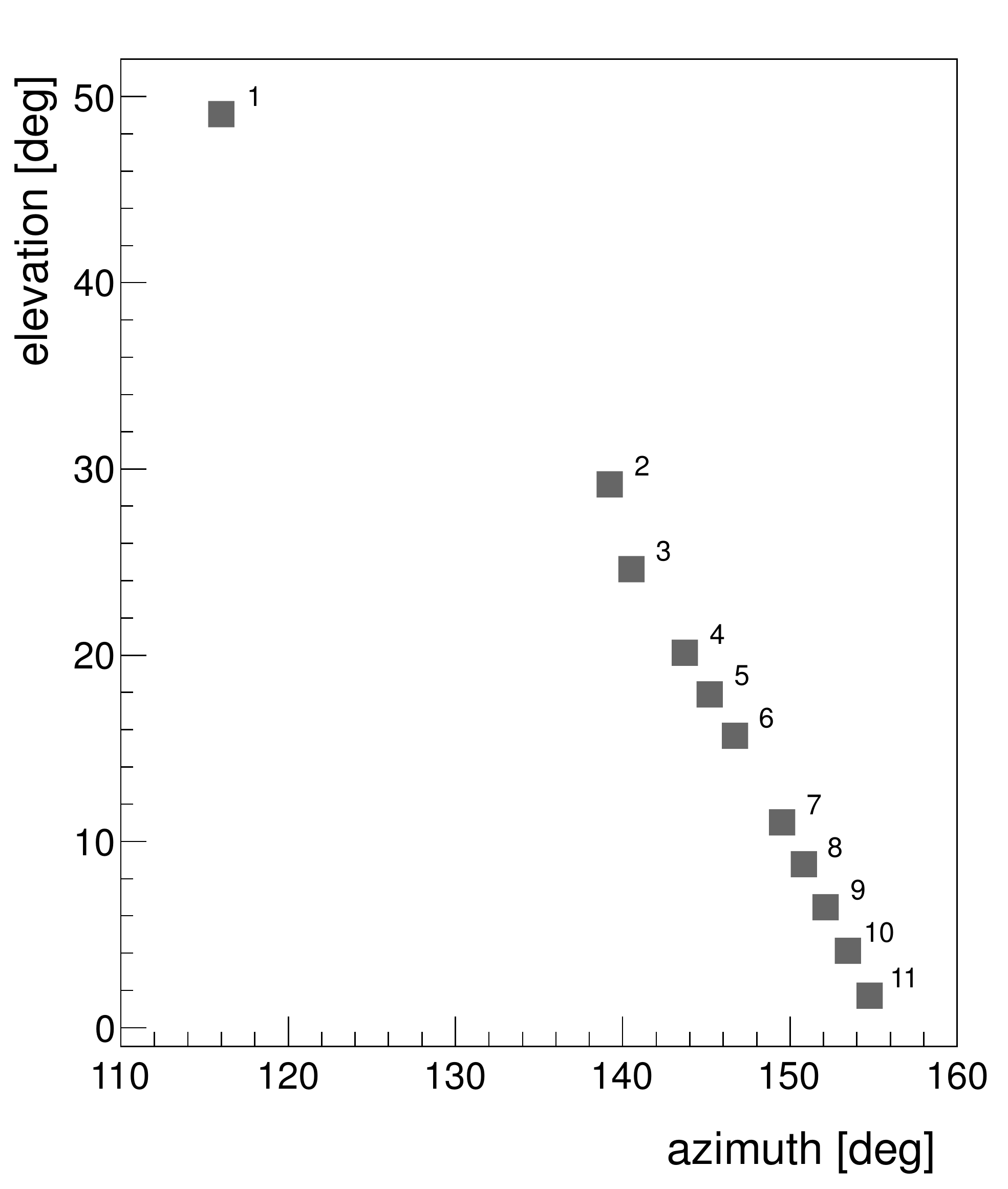}
    \caption{\label{fig:fram_cloud_images} \slshape
      \textsl{Left}: A series of eleven wide-field (WF) CCD images recorded
      during the sampling of a shower track.
      \textsl{Right}: the alt-azimuth coordinates of the centers of each of the
      eleven images shown in the left panel.
      Note that the images have not been flat-fielded, causing the image
      borders to appear darker than the centers.  Also note that the WF camera
      is tilted about $20^\circ$ from vertical (see image 11, where the Andes
      Mountains fill the bottom half of the image).  In images 5-7, an
      optically thick cloud is observed to block the stars.
      }
  \end{center}
\end{figure}

\begin{figure}[ht]
  \begin{center}
    \includegraphics*[width=.52\linewidth,clip,angle=-90]{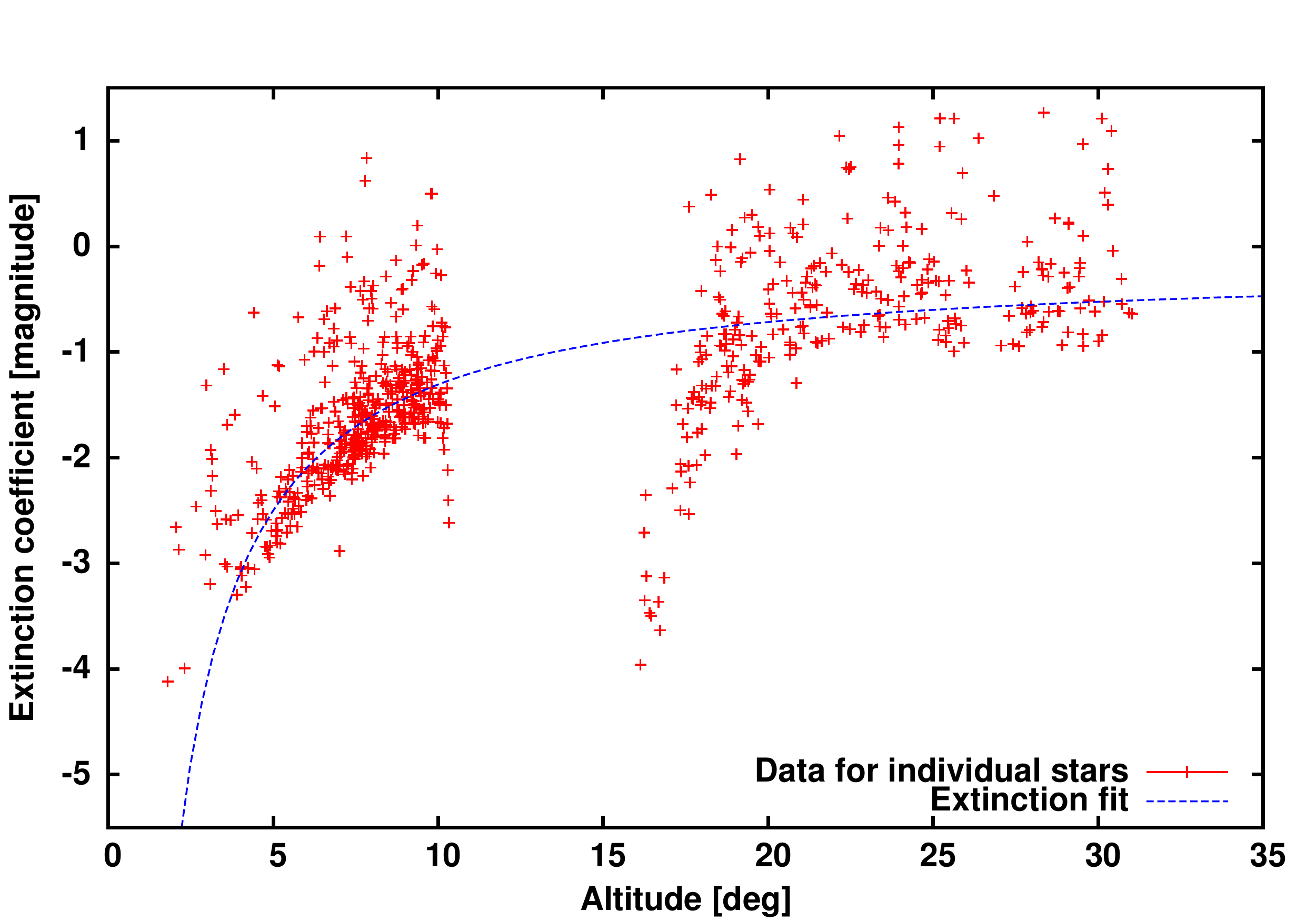}
    \caption[]{\label{fig:fram_result} \slshape
      The analysis of the series of images shown in
      Fig.~\ref{fig:fram_cloud_images}. The computed extinctions for the
      individual stars are shown along the extinction fit for the clear sky.
      The prominent drop between $10^{\circ}$ and $17^{\circ}$ clearly
      indicates the presence of the optically thick cloud. 
    }
  \end{center}
\end{figure}

The output of the FRAM rapid monitoring analysis is a set of extinction
coefficients $\{k\}$ computed in the Johnson-Cousins B filter for each star
detected in the WF camera image.  A complete shower track is covered by
10--20~WF CCD images (Fig.~\ref{fig:fram_cloud_images}), and so we obtain
several hundred to several thousand stars with computed extinction coefficients
per track (Fig.~\ref{fig:fram_result}).

To find obscurations along the line of sight to the shower, we compare the
extinction data to a ``reference'' profile of the extinction in a clear sky.
The clear-sky profile is estimated by assuming a uniform atmosphere in which the
extinction depends only on the airmass.  This gives a zenith-dependent reference
curve which can be compared to the extinction coefficients determined using the
WF images.  For example, the extinction coefficients in
Fig.~\ref{fig:fram_result} show a dramatic deviation from the clear-sky
reference between $10^\circ$ and $15^\circ$ elevation due to the presence of the
clouds visible in Fig.~\ref{fig:fram_cloud_images}.  

\subsubsection{Air Shower Analysis using FRAM rapid monitoring data
  \label{subsec:reco_fram}}

To determine whether or not the FRAM observations can be used to identify
prominent changes in the extinction which affect the air shower reconstruction,
we have searched the 242 events monitored with FRAM in 2010.  Within this
sample, 13 extinction profiles indicated the presence of clouds. 

After inspection of the corresponding longitudinal profiles recorded with the FD
telescopes, we have found that 7 of the 13 events also passed the standard
quality filter, which includes a tight cut on the profile reconstruction
($\chi^2_\text{GH}/\text{n}_\text{dof}<2.5$) that tends to remove events
affected by clouds.  Hence, out of 242 showers which satisfy all standard
criteria for high-quality hybrid observations, 7 events (3\%) were clearly
recorded with clouds in the shower-detector plane.

We note at this point that there are further cuts applied on data for physics
analyses, e.g., mass composition studies or building an energy spectrum, which
remove events affected by clouds.  The most important requires the cloud
coverage in the sky above each telescope to be below 20\% during a given
measurement hour.  The cloud coverage is estimated from the lidar
\emph{AutoScan}, in which the lidar stations continuously sweep the high-zenith
regions above the FD telescopes.  Using the 20\% cloud coverage cut, we find
that 10 of the 13 cloud-affected FRAM observations will be removed from physics
analyses.  Of the remaining three events, two were observed during periods of
broken cloud coverage with non-zero coverage in the neighboring measurement
hours.  However, one event was observed during a night in which the lidar data
indicates cloud-free conditions during the whole night.  After applying the cut
on the profile reconstruction quality, only one of these three events passed the
standard criterion.

It may appear surprising that the lidar analysis could miss a prominent cloud
layer clearly observed with the FRAM telescope, but we must note one critical
limitation of the FRAM analysis.  Because the extinction is determined using
stellar magnitudes, the FRAM measurement only describes the overall atmospheric
influence along the line of sight between the FD telescopes and the top of the
atmosphere.  Using the FRAM alone, it is not possible to determine whether or
not a shower profile was observed behind a cloud or in the foreground.  In order
to determine whether or not a cloud seen by FRAM actually affected a
fluorescence profile, it is necessary to look into the FD data themselves to see
if the profiles are affected by spurious gaps (absorption) or spikes (multiple
scattering).  Of the 13 cloud-affected events discussed above, inspection of the
data suggests that 11 showers originated behind or partially inside the observed
clouds.

This study illustrates the trade-off between passive and active atmospheric
monitoring for the FD calibration; the FRAM can operate during FD DAQ without
constraints, but the resulting measurements do not have the very useful slant
depth resolution of the lidar data.

Nevertheless, the rapid monitoring observations performed with FRAM confirm that
the standard selection of hybrid events for physics analysis at the \pao is
efficiently removing profiles affected by atmospheric scattering.  The
contamination of the high-quality standard data set by the presence of clouds
was found to be of order 1\%.  The problematic cases within this subsample will
be further studied in order to improve future data processing.

\subsubsection{Analysis of the double-bump events observed by the FRAM telescope
\label{subsec:dbtrigfram}}

Because of the short operation of the double-bump rapid monitoring observations,
only a few candidates were analyzed for this purpose, and all of them are
compatible with profiles containing clouds.  As with \sts, the anomalous profile
trigger will need to be tuned using cloudy and clear data before it can become
truly effective.

\section{Conclusion
\label{sec:conclusion}}

A rapid atmospheric monitoring program has been implemented and operated
successfully at the \pao since early 2009.  The online reconstruction has
produced reliable data which have been used to trigger dedicated observations
of the atmosphere.  Among the showers of particular interest are those
initiated by very high-energy cosmic rays, or those with unusual longitudinal
profiles.  Since the start of rapid monitoring, there have been $52$~dedicated
balloon flights (until the termination of the balloon program end of December
2010), $112$~lidar scans (through October 2011), and $586$~observations with an
optical telescope (through July 2011).  The program has demonstrated that it is
possible to perform targeted atmospheric monitoring based on real-time cosmic
ray measurements.

The \emph{Balloon-the-Shower} program was a useful tool for studying short-term
variations in the altitude-dependent profiles of the main atmospheric state
variables. Using actual atmospheric profiles instead of monthly models in
dedicated air shower analyses improves the accuracy of the \Offline
reconstruction.  Unfortunately, this part of the rapid monitoring program
imposed a large burden on the collaboration, and due to technical problems the
observations fell short of the expected rate.  For these reasons, the program
was discontinued at the end of 2010.  However, the measurements have had a
positive effect, as we have used \bts measurements to confirm the validity of
atmospheric profiles from the Global Data Assimilation System (GDAS).  Profiles
from GDAS, which have more regularity and better time resolution than the
balloon flights, are now part of the standard \Offline reconstruction.

For the \emph{Shoot-the-Shower} program, we have configured the lidar
telescopes to search for showers affected by clouds during periods that might
otherwise be flagged as cloud-free by the standard hourly lidar scans.  A
similar program has been implemented using the FRAM optical telescope, but
without the operational limitations imposed on the lidar.  Analysis of the \sts
and FRAM measurements indicate that it is relatively difficult to find events
affected by atmospheric anomalies during periods that are identified as clear
by the standard monitoring program.  In other words, the rapid monitoring
confirms that existing cuts on the quality of the shower profile reconstruction
and on atmospheric conditions are effective at removing showers distorted by
clouds and aerosol layers.

Since the \sts and FRAM measurements require much less human intervention than
\bts, we expect that these programs will continue for at least several more
years.  Both programs have recently incorporated a trigger based on the search
for anomalous double-bump shower profiles.  Analyses of the early data
indicate that some real-time restrictions of double-bump monitoring --
especially during periods of high cloud coverage -- are necessary to make this
program effective.  Real-time cloud monitoring should be possible using the
standard lidar scans.  A trigger filter based on total cloud coverage is
currently under development.

Finally, we have found that the rapid monitoring program can be extended easily
to incorporate other instruments.  It is particularly useful for measurements
which must be restricted to avoid interference with the fluorescence
telescopes.  In 2012, a Raman lidar will be installed at the center of the \pao,
and we expect that this instrument will become part of the rapid monitoring
program.

\section*{Acknowledgments}
The successful installation, commissioning, and operation of the Pierre Auger Observatory
would not have been possible without the strong commitment and effort
from the technical and administrative staff in Malarg\"ue.

We are very grateful to the following agencies and organizations for financial support: 
Comis\-i\'on Nacional de Ener\-g\'ia At\'omica, 
Fundaci\'on Antorchas,
Gobierno De La Provincia de Mendoza, 
Municipalidad de Malarg\"ue,
NDM Holdings and Valle Las Le\~nas, in gratitude for their continuing
cooperation over land access, Argentina; 
the Australian Research Council;
Conselho Nacional de Desenvolvimento Cient\'ifico e Tecnol\'ogico (CNPq),
Financiadora de Estudos e Projetos (FINEP),
Funda\c{c}\~ao de Amparo \`a Pesquisa do Estado de Rio de Janeiro (FAPERJ),
Funda\c{c}\~ao de Amparo \`a Pesquisa do Estado de S\~ao Paulo (FAPESP),
Minist\'erio de Ci\^{e}ncia e Tecnologia (MCT), Brazil;
AVCR AV0Z10100502 and AV0Z10100522, GAAV KJB100100904, MSMT-CR LA08016,
LG11044, MEB111003, MSM0021620859, LA08015 and TACR TA01010517, Czech Republic;
Centre de Calcul IN2P3/CNRS, 
Centre National de la Recherche Scientifique (CNRS),
Conseil R\'egional Ile-de-France,
D\'epartement  Physique Nucl\'eaire et Corpusculaire (PNC-IN2P3/CNRS),
D\'epartement Sciences de l'Univers (SDU-INSU/CNRS), France;
Bundesministerium f\"ur Bildung und Forschung (BMBF),
Deutsche Forschungsgemeinschaft (DFG),
Finanzministerium Baden-W\"urttem\-berg,
Helmholtz-Gemeinschaft Deutscher Forschungszentren (HGF),
Ministerium f\"ur Wissen\-schaft und For\-schung, Nordrhein-Westfalen,
Ministerium f\"ur Wissen\-schaft, For\-schung und Kunst, Baden-W\"urttem\-berg, Germany; 
Istituto Nazionale di Fisica Nucleare (INFN),
Ministero dell'Istruzione, dell'Universit\`a e della Ricerca (MIUR), Italy;
Consejo Nacional de Ciencia y Tecnolog\'ia (CONACYT), Mexico;
Ministerie van Onderwijs, Cultuur en Wetenschap,
Nederlandse Organisatie voor Wetenschappelijk Onderzoek (NWO),
Stichting voor Fundamenteel Onderzoek der Materie (FOM), Netherlands;
Ministry of Science and Higher Education,
Grant Nos. N N202 200239 and N N202 207238, Poland;
Portuguese national funds and FEDER funds within COMPETE - Programa Operacional Factores de Competitividade through 
Funda\c{c}\~ao para a Ci\^{e}ncia e a Tecnologia, Portugal;
Ministry of Education, Science, Culture and Sport,
Slovenian Research Agency, Slovenia;
Comunidad de Madrid, 
Consejer\'ia de Educaci\'on de la Comunidad de Castilla La Mancha, 
FEDER funds, 
Ministerio de Ciencia e Innovaci\'on and Consolider-Ingenio 2010 (CPAN),
Xunta de Galicia, Spain;
Science and Technology Facilities Council, United Kingdom;
Department of Energy, Contract Nos. DE-AC02-07CH11359, DE-FR02-04ER41300,
National Science Foundation, Grant No. 0450696,
The Grainger Foundation USA; 
NAFOSTED, Vietnam;
Marie Curie-IRSES/EPLANET, European Particle Physics Latin American Network, 
European Union 7th Framework Program, Grant No. PIRSES-2009-GA-246806; 
European Union 6th Framework Program, Grant No. MEIF-CT-2005-025057; 
European Union 7th Framework Program, Grant No. PIEF-GA-2008-220240; 
and UNESCO.

\bibliographystyle{JHEP}
\bibliography{xts_paper}

\clearpage

\end{document}